\documentclass[journal]{IEEEtran} 
\ifCLASSINFOpdf
\else
\fi

\usepackage{cite}
\usepackage{stfloats}
\ifCLASSINFOpdf
   \usepackage[pdftex]{graphicx}
   \graphicspath{{../pdf/}{../jpeg/}}
   \DeclareGraphicsExtensions{.pdf,.jpeg,.png}
\else
   \usepackage[dvips]{graphicx}
   \graphicspath{{../eps/}}
   \DeclareGraphicsExtensions{.eps}
\fi

\usepackage[cmex10]{amsmath}
\usepackage{mathrsfs,color,amssymb}
\usepackage{amsfonts}

\usepackage{algorithm}
\usepackage{algorithmic}
\allowdisplaybreaks

\ifCLASSOPTIONcompsoc
\else
  \usepackage[caption=false,font=footnotesize]{subfig}
\fi

\usepackage{hyperref}
\usepackage{graphicx}
\usepackage{epstopdf}

\newtheorem{theorem}{\bf{Theorem}}
\newtheorem{lemma}{\bf{Lemma}}
\newtheorem{corollary}{\bf{Corollary}}
\newtheorem{proposition}{\bf{Proposition}}

\newtheorem{remark}{\textit{Remark}}
\newtheorem{definition}{\bf{Definition}}

\begin{document}

%

\title{Performance Scaling Law for Multi-Cell Multi-User Massive MIMO}
%
%
%

\author{Cheng~Zhang,~\IEEEmembership{Student Member,~IEEE,}
        Yindi~Jing,~\IEEEmembership{Member,~IEEE,}
        Yongming~Huang,~\IEEEmembership{Member,~IEEE,}

        and Luxi Yang,~\IEEEmembership{Member,~IEEE}
\thanks{
Copyright (c) 2015 IEEE. Personal use of this material is permitted. However, permission to use this material for any other purposes must be obtained from the IEEE by sending a request to pubs-permissions@ieee.org.

C. Zhang, Y. Huang, and L. Yang are with the School of Information Science and Engineering, Southeast University, Nanjing 210096, P. R. China (email: {zhangcheng1988, huangym, lxyang}@seu.edu.cn).

Y. Jing is with the Department of Electrical and Computer Engineering, University of Alberta, Edmonton, Canada, T6G 1H9 (email: yindi@ualberta.ca).
}}

\maketitle

\begin{abstract}
This work provides a comprehensive scaling law based performance analysis for multi-cell multi-user massive multiple-input-multiple-output (MIMO) downlink systems. Imperfect channel state information (CSI), pilot contamination, and channel spatial correlation are all considered. First, a sum-rate lower bound is derived by exploiting the asymptotically deterministic property of the received signal power, while keeping the random nature of other components in the signal-to-interference-plus-noise-ratio (SINR) intact. Via a general scaling model on important network parameters, including the number of users, the channel training energy and the data transmission power, with respect to the number of base station antennas, the asymptotic scaling law of the effective SINR is obtained, which reveals quantitatively the tradeoff of the network parameters. More importantly, pilot contamination and pilot contamination elimination (PCE) are considered in the analytical framework. In addition, the applicability of the derived asymptotic scaling law in practical systems with large but finite antenna numbers are discussed. Finally, sufficient conditions on the parameter scalings for the SINR to be asymptotically deterministic in the sense of mean square convergence are provided, which covers existing results on such analysis as special cases and shows the effect of PCE explicitly.
\end{abstract}


%
\IEEEpeerreviewmaketitle

\section{Introduction}
Massive multiple-input multiple-output (MIMO) (also known as large-scale MIMO) is considered as one of the most promising technologies for next generation wireless communication systems \cite{Label9,Label8}, since it can provide very high spectral and energy efficiencies. Besides, for massive MIMO systems, the effects of small-scale fading and interference can be significantly reduced with linear signal processing, such as maximal-ratio transmission (MRT) and zero-forcing (ZF) \cite{Label10}.

The performance of massive MIMO single/multi-cell downlink systems has been widely studied in the literature. In \cite{Label11}, the asymptotic signal-to-interference-plus-noise-ratio (SINR) for linear precoding in single-cell \textit{frequency-division-duplex} (FDD) systems was derived based on which the optimization of the user number was investigated. In \cite{Label1}, the power scaling, i.e., how fast the transmission power can decrease with the antenna number while maintaining a certain SINR level, for multi-cell \textit{time-division-duplex} (TDD) systems was given based on asymptotic SINR analysis, where the pilot contamination was shown to be the only performance limit. In \cite{Label4}, capacity lower bounds and energy efficiency of MRT and ZF in single-cell TDD systems were studied and compared with each other. In the parallel field of uplink systems, similar results were derived in \cite{Label2}, \cite{Label3}. These works mainly focus on the derivations of asymptotically deterministic expressions for the SINR, and the performance analysis only consider some special cases, e.g., the constant or linearly increasing user number with respect to the base station (BS) antenna number, the power scaling for perfect or imperfect channel state information (CSI) with equal training and transmission power and constant user number. A mathematical counterpart can be seen in randomly spread CDMA \cite{Sedaghat_CDMA} where the optimum asymptotic multiuser efficiency can hold when the ratio of the user number to the chip number is kept constant or grows logarithmically with the user number.

Recently, along this route, the performance improvement for using coherent detection based on downlink channel estimation was shown in \cite{Label5,Label28,Label29}, especially for large but finite antenna numbers. Meanwhile, the impact of other practical factors such as channel aging \cite{Label14}, \hspace{-5pt} \cite{Kong_sumrate}, hardware limits \cite{Label15}, and Rician fading \cite{Label16} on the performance of massive MIMO systems was studied. However, a comprehensive scaling law analysis for multi-cell multi-user massive MIMO is still missing in existing works.


Another important question in the performance analysis of massive MIMO is when the derived asymptotic results are applicable in practical systems. While many works derived deterministic SINR approximations via calculating the expected value of the SINR or its compenents, the analysis on whether the instantaneous SINR becomes asymptotically deterministic and converges to the derived expressions is sometimes missing.
In \cite{Label1}, this problem was studied through the so called massive MIMO efficiency for the ideal situation that the noise, channel estimation error, and interference are negligible compared with pilot contamination, where the dependence of the massive MIMO efficiency on the ratio of spatial degrees of freedom/antenna number to the user number and the product of antenna number and transmit power was shown by simulations. In \cite{Label21}, via simulations, the convergence speed of the strong law of large numbers with respect to the antenna number and the convergence of the SINR of MRT and ZF to their expected values were both analyzed for fixed ratio of antenna number to user number. It was shown that the SINR of MRT has slower convergence than that of ZF, while the explicit relationship between the convergence speed and important parameters, i.e., antenna number, user number, transmit power, and channel training energy (CSI quality), is missing.

In this work, by drawing lessons from the analytical method and the mean square convergence definition in \cite{Label33}, we provide a comprehensive performance scaling law analysis for multi-cell multi-user TDD massive MIMO downlink systems with channel spatial correlation, CSI error, and pilot contamination, and investigate the applicability of the derived asymptotic scaling law in practical systems with large but finite antenna numbers. Extensive simulations are also conducted to validate the analytical results. Our main new contributions are summarized as follows.
\begin{itemize}
\item We consider a multi-cell multi-user massive MIMO network with correlated channel model, CSI error, and pilot contamination. In dealing with pilot contamination, recent advancements in pilot contamination elimination (PCE)\cite{Muller_Blind,Label12,Label34,Label35,Label13} are also considered, while in existing works it is treated as a constant bottleneck of massive MIMO. An approximate model for the PCE effect is proposed.
\item A lower bound on the system sum-rate is derived based on which a general performance scaling law is obtained. The result reveals the effect and joint interaction of extensive network parameters, i.e, the channel training energy, the transmission power, and the user number. In the scaling-law analysis, a general model is used where  the parameters have continuous scaling exponents with respect to the number of BS antennas, while in most existing work, only a few discrete values for the power scaling exponent, e.g., $0$, $1$, $1/2$, are allowed.

\item To understand the applicability of the derived scaling law in practical systems with large but finite antenna number, the effect of system parameters on its accuracy is analyzed and evaluated by simulations. 
\item We conduct quantitative analysis on the asymptotically deterministic property of random instantaneous SINR using the mean square convergence. Sufficient conditions for the instantaneous SINR to be asymptotically deterministic are given, from which the tradeoff among different parameter scales is discovered.
\end{itemize}

The major differences of our work compared with \cite{Label28}, \cite{Label29}, and \cite{Label33} are summarized as follows.
\begin{itemize}
\item The references \cite{Label28, Label29} focus on the effect of downlink CSI training on the performance of multi-cell multi-user massive MIMO TDD networks with uncorrelated channels via incorporating the the minimum-mean-squared-error (MMSE) downlink training into the performance derivations. Whereas, we focus on the performance scaling law analysis and the asymptotic convergence property of the SINR in multi-cell multi-user massive MIMO TDD networks with correlated channels. Meanwhile, in this work, a sum-rate lower bound is derived based on quantitative analysis with mean square convergence while the same sum-rate expression is derived as an approximation in \cite{Label28, Label29} based on the law of large numbers.
\item The reference \cite{Label33} provides the performance scaling law and the SINR convergence analysis for multi-user massive MIMO relay networks with MRT/MRC operation. While this work adopts the mean square convergence idea and the SINR analysis method from \cite{Label33}, it considers  multi-cell multi-user massive MIMO downlink networks without relaying. Also, \cite{Label33} is on independent and identically distributed (i.i.d.) channels only while this work adopts a general spatially correlated channel model. Thus, it is fundamentally different to \cite{Label33} in both network model and channel model. Meanwhile, both MRT and ZF precoders are studied in this work. Furthermore, the pilot contamination and PCE are taken into account in our work due to the multi-cell systems. The corresponding derivation procedures and analytical results are significantly different. Finally, the analysis on the effectiveness of the derived scaling law for practically not-so-large antenna numbers is completely new.
\end{itemize}

The remaining of the paper is organized as follows. In Section II, the channel model and system model including both the multi-cell channel estimation with pilot contamination and the downlink transmission with MRT are introduced. Section III shows the performance scaling law result, discussions on typical network scenarios, and the analysis on the applicability to practical systems with large but finite antenna numbers. In Section IV, sufficient conditions for the SINR to be asymptotically deterministic are derived. In Section V, the ZF counterpart is provided. Section VI shows simulations and conclusions are provided in Section VII.

In this paper, bold upper case letters and bold lower case letters are used to denote matrices and vectors, respectively. For a matrix $\bf A$, its conjugate transpose, transpose, and trace are denoted by ${\bf A}^H$, ${\bf A}^T$ and ${\rm tr}({\bf A})$, respectively. ${\rm E}(\cdot)$ and ${\rm Var}(\cdot)$ are the mean and variance operators. $\cong $ means equivalence in probabilistic distribution. $W_p(n, {\bf V})$ denotes the Wishart distribution with degrees of freedom $n$ and scale matrix ${\bf V}\in\mathbb{C}^{p\times p}$. The notation ${a} = \mathcal{O}\left( {{b}} \right)$ means that when $M \to \infty $, ${a}$ and ${b}$ have the same scaling with respect to $M$, in other words, for large $M$, there exists positive constants $c_1$ and $c_2$, such that $c_1\le |a/b|\le c_2$.

\section{System Model}
Consider a multi-cell multi-user massive MIMO network\footnote{We  use massive MIMO networks to denote multi-cell multi-user massive MIMO networks in this paper unless otherwise mentioned.} consisting of $L > 1$ cells with one $M$-antennas BS and $K$ scheduled single-antenna users in each cell. The BSs and users are assumed to operate a TDD protocol with universal frequency reuse.
\subsection{Channel Model}
The uplink spatially correlated channel from User $m$ in Cell $j$ to the antennas of BS $l$ in Cell $l$ can be written as
\begin{equation}\label{Eq2}
{{\bf{h}}_{ljm}} = {\bf{R}}_{ljm}^{{1 \mathord{\left/
 {\vphantom {1 2}} \right.
 \kern-\nulldelimiterspace} 2}}{{\bf{z}}_{ljm}},
\end{equation}
where ${\bf R}_{ljm}$ is the channel correlation matrix and ${\bf z}_{ljm}$ is the independent fast-fading channel vector. This channel model is widely adopted in massive MIMO literature and applies to systems with reasonable antenna array size and BS-users distance. However, such a correlation-based stochastic model actually ignores spherical wavefront and non-stationarity in the spatial domain \cite{Wang_recent,Wu_non,Wu_Stationary}.

The following assumptions on ${\bf R}_{ljm}$ are made by taking into consideration the pilot contamination, PCE, and tractability analysis. The set $\mathbb{S}_l^{p,s}$ denotes the index set of all cells having pilot contamination to Cell $l$ while $\mathbb{S}_l^{o,s}$ denotes the index set of remaining cells whose pilot contamination is eliminated. We assume that for all $l,m$ and $j \in \{l\} \cup \mathbb{S}_l^{p,s}$,
\begin{equation}\label{Eq4}
{\bf{R}}_{ljm} = \beta_{ljm}{\bf{A}}{\bf{A}}^H,
\end{equation}
where the channel direction matrix ${\bf A}$ is an ${M \times \Delta}$ unitary matrix and $\Delta=cM$ with $c\in(0,1]$ indicates the channel correlation level. For example, elements of the channel vector become i.i.d.~when $c = 1$. Also, $\beta_{ljm}$ is the pass-loss fading factor where $\beta_{llm}=M/\Delta$ and $\beta_{ljm}=\alpha M/\Delta$. The parameter $\alpha\in(0,1)$ denotes the large-scale fading between different cells and the intra-cell one is normalized to one\footnote{In this work, for tractable analysis and to focus on the scaling law analysis, the parameter $\alpha$ is used to represent the large-scale fading between different cells. Strictly speaking, the parameter depends on the user locations. But this simplification will not affect the scaling law analysis result in Section III-B for constant path-loss exponent and finite distances. It may affect the details of conclusions on the applicability of the scaling law for not-so-large $M$ in Section III-D. Further, for the effect of $\alpha$ on the sum-rate performance, please refer to \cite[Fig. 3]{Label28}.}.
For $j \in \mathbb{S}_l^{o,s}$, the pilot contamination from Cell $j$ to Cell $l$ is eliminated by using the spatial information
in channel correlation matrix. Thus we have
\begin{equation}\label{Eq5}
{{\bf{A}}^H}{{\bf{R}}_{ljm}} = {\bf{0}}.
\end{equation}


This channel model considers both the channel correlation and the inter-cell pass-loss fading. It is similar to the one used in \cite[Section III]{Label1} where motivations and justifications for this model can be referred to. A new feature of the modeling in \eqref{Eq4} and \eqref{Eq5} is that the possible effect of PCE, i.e., the BSs schedule users with relatively less pilot contamination for channel training and transmission, can be embodied. In \cite{Label12,Label34,Label35,Label13}, several efficient methods by exploiting the statistical channel information have been proposed to alleviate the pilot contamination\cite{Label1}, \cite{Label23}. Generally speaking, the overlap among users' channel directions can be in any degree leading to different pilot contamination level \cite{Label12}. For explicit analytical results, in \eqref{Eq4}, we assume that channels from both users in Cell $l$ and users in other cells with pilot contamination to Cell $l$ have the same $\bf A$ matrix, while in other cells, pilot contamination is eliminated by having orthogonal channel directions as shown in \eqref{Eq5}. This is a simplified discretization approximation of the PCE effect and the extension to the case of more general channel direction distribution is left for future research.

\subsection{Uplink Channel Estimation for Precoding}
In the uplink training phase, users in Cell $j$ transmit mutually orthogonal pilot sequences which are reused by users in other cells. Denote the length of the pilot sequence as $\tau$ ($\tau \ge K$ for reliable performance \cite{Label20}). Assume that all $K$ users in each cell use the same training power $P_t$. Define $E_t\triangleq \tau P_t$ which is the channel training energy. By correlating the received training signals with the pilot sequence of User $k$, BS $j$ has the observation ${\bf y}_{jk}^{tr}\in\mathbb{C}^M$ as
\begin{equation}\label{Eq6}
{\bf y}_{jk}^{tr}=\sqrt{E_t}{\bf h}_{jjk}+\sqrt{E_t}\sum_{l\ne j}{\bf h}_{jlk} +{\bf n}_{jk}^{tr},
\end{equation}
where ${\bf n}_{jk}^{tr}\sim \mathcal{CN}({\bf 0},{\bf I}_M)$ is the local noise. With MMSE estimation \cite{Label20}, the estimate ${\bf \hat h}_{jlk}$ is given as
\setlength{\arraycolsep}{1pt}
\begin{eqnarray}
{\bf \hat h}_{jlk}&=& \sqrt{E_t}{\bf R}_{jlk}{\rm E}\{{\bf y}_{jk}^{tr}{\bf y}_{jk}^{tr,H}\}^{-1}{\bf y}_{jk}^{tr} \nonumber\\
&=&\sqrt{E_t}{\bf R}_{jlk}\left(\hspace{-1mm}E_t{\bf R}_{jjk}+E_t\sum_{l\ne j}{\bf R}_{jlk}+{\bf I}_M \hspace{-1mm}\right)^{\hspace{-1mm}-1}\hspace{-2mm}{\bf y}_{jk}^{tr}.
\label{Eq7}
\end{eqnarray}
\setlength{\arraycolsep}{5pt}We assume that all the $L$ cells have the same pilot contamination level, i.e., $\left| {{{\mathbb{S}}}_j^{p,s}} \right| =  {{{L}_{p}}} \in[0,L-1]$ for all $j$.
By defining
\begin{equation}\label{Eq8}
{Q} = {\left( {\frac{c}{{{E_t}}} + 1 + {L}_{p}\alpha } \right)^{ - 1}}
\end{equation}
and utilizing \eqref{Eq4}, \eqref{Eq5}, we have ${\bf \hat h}_{jlk}\sim\mathcal{CN}({\bf 0},{\bf \hat \Theta}_{jlk})$ for $l \in \{j\}\cup{{\mathbb{S}}}_j^{p,s}$ where
\begin{equation}\label{Eq9}
{{\bf{\hat\Theta }}_{jlk}} =cQ\beta_{jlk}^2{\bf{A}}{{\bf{A}}^H}.
\end{equation}
From \eqref{Eq7}, it can be shown that
\begin{equation}\label{Eq11}
{\bf \hat h}_{jlk}=c\beta_{jlk}{\bf \hat h}_{jjk}\ \text{ for }\ l \in \{j\}\cup{{\mathbb{S}}}_j^{p,s}.
\end{equation}
This will be utilized to simplify the derivations in later parts.

Define ${\bf  {\tilde h}}_{jlk}={\bf h}_{jlk}-{\bf \hat h}_{jlk}$, which is the estimation error. Due to the feature of MMSE estimation and Gaussian distribution, ${\bf{\hat h}}_{jlk}$ and ${\bf {\tilde h}}_{jlk}$ are mutual independent and  ${\bf{\tilde h}}_{jlk}\sim\mathcal{CN}({\bf 0},{\bf }{\bf \tilde\Theta}_{jlk})$ where
\begin{equation}\label{Eq10}
{{\bf{\tilde \Theta }}_{jlk}} =\beta_{jlk}(1-cQ\beta_{jlk}){\bf{A}}{\bf{A}}^H.
\end{equation}
Note that $Q$ denotes the CSI quality, which is dependent on both channel training energy and pilot contamination. When $E_t \to \infty$, we have $Q\to{\left( {1+{L}_{p}\alpha } \right)^{ - 1}}$ and the CSI quality is only affected by pilot contamination. If further perfect pilot contamination elimination is achieved, we have ${L}_{p}=0$, $Q\to1$ and perfect CSI can be obtained.


\subsection{Downlink Transmission with MRT}
MRT precoding has low computational complexity, robustness, and high asymptotic performance \cite{Label9}. With the channel estimation in (\ref{Eq7}), the MRT precoding matrix of BS $l$ is ${\bf \hat H}_{ll}=[{\bf \hat h}_{ll1},...,{\bf \hat h}_{llK}]\in\mathbb{C}^{M \times K}$. By assuming  channel reciprocity, the received signal $y_{jm}$ of User $m$ in Cell $j$ is given as
\begin{equation}\label{Eq1}
y_{jm}=\sqrt{\frac{\rho}{KMQ}}\sum_{l=1}^{L}{\bf h}_{ljm}^{H}{{\bf \hat H}_{ll}{\bf x}_l}+n_{jm},
\end{equation}
where $\rho$ denotes the average downlink transmission power or the signal-to-noise-ratio (SNR), 
${\bf x}_l=[x_{l1},...,x_{lK}]^T\in\mathbb{C}^{K}\sim\mathcal{CN}({\bf0},{\bf I}_K)$ contains the data symbols for the $K$ users in Cell $l$, and $n_{jm}\sim \mathcal{C}\mathcal{N}(0,1)$ is the receiver noise. 

Further, $y_{jm}$ can be rewritten as
\begin{equation}\label{Eq12}
\begin{array}{c}
\hspace{0mm}y_{jm}
=\underbrace {{\sqrt{\hspace{-1mm}\frac{\rho}{KMQ}}{\bf \hat h}_{jjm}^H{\bf \hat h}_{jjm}}x_{jm}}_{\rm Desired \ signal}
+\hspace{-1mm}\underbrace {{\sqrt{\hspace{-1mm}\frac{\rho}{KMQ}}{\bf \hat h}_{jjm}^H \hspace{-1mm}\sum_{k\ne m}\hspace{-0.5mm}{\bf \hat h}_{jjk}}x_{jk}}_{\rm Intra-cell \ interference  }\\
\hspace{-2.5cm}+\underbrace {{{\sqrt{\hspace{-1mm}\frac{\rho}{KMQ}}\sum_{l\ne j}\sum_{k=1}^{K}{\bf \hat h}_{ljm}^H{\bf \hat h}_{llk}x_{lk}}}}_{\rm Inter-cell \ interference} \\
\hspace{-15mm}+ \underbrace {{\sqrt{\hspace{-1mm}\frac{\rho}{KMQ}}\sum_{l=1}^{L}\sum_{k=1}^{K}{\bf \tilde h}_{ljm}^H{\bf \hat h}_{llk}}x_{lk}}_{\rm Noise  \ due\ to \ CSI\ error} + n_{jm}.
\end{array}
\end{equation}In this formulation, the effective channels ${\bf \hat{h}}_{ljm}^H{\bf \hat h}_{llk}, \forall l,k$ are assumed to be known at User $m$ in Cell $j$ for coherent detection. This assumption has been used in recent works on TDD massive MIMO relay networks \cite{Label32,Label26,Label27}. One possible method to obtain this information is to have BS $l$ send the information of ${\bf \hat{h}}_{ljm}^H{\bf \hat h}_{llk}, \forall k$ to User $m$ in Cell $j$ after the uplink training period. This is similar to the feedback procedure in FDD systems where the users feedback their CSI to the associated BSs. Another method is to obtain effective downlink channels via downlink training \cite{Label5,Label28,Label29}. In this case, an estimate of ${\bf h}_{ljm}^H{\bf \hat h}_{llk }, \forall l,k$ can be obtained. Strictly speaking, in this case, the estimate instead of ${\bf \hat h}_{ljm}^H{\bf \hat h}_{llk }$ should be used in the analysis. But it actually dose not affect the sum-rate performance for large $M$ (Details are omitted here). Therefore, our assumption can also be interpreted as using ${\bf \hat h}_{ljm}^H{\bf \hat h}_{llk }$ as an approximation of the estimate of ${\bf h}_{ljm}^H{\bf \hat h}_{llk}$ to avoid insignificant details of the downlink training to better focus on the main purpose of this paper without loss of accuracy.

Since the BS contaminated by one user creates interference to the user in the downlink transmission, we define ${{\mathbb{S}}}_j^{p,d} \buildrel \Delta \over = \left\{ {l\left| {l \ne j,j \in {{\mathbb{S}}}_l^{p,s}} \right.} \right\}$ and ${{\mathbb{S}}}_j^{o,d} \buildrel \Delta \over = \left\{ {l\left| {l \ne j,j \in {{\mathbb{S}}}_l^{o,s}} \right.} \right\}$ which are the index sets of cells or BSs which are contaminated and not contaminated by users in Cell $j$ in the uplink training period, respectively. For simplicity, we assume $\left| {{{\mathbb{S}}}_j^{p,d}} \right| = \left| {{{\mathbb{S}}}_j^{p,s}} \right|= {L}_{p} $ for all $j$.

Define
{\small
\begin{align}
\label{Eq13}& {P_{s}} = \frac{{\left| {{\bf \hat h}_{jjm}^H{\bf \hat h}_{jjm}} \right|^{2}}}{{{M^2}}},\\
\label{Eq14}&{P_{i,in}} =  \frac{1}{{K - 1}}\sum\limits_{k \ne m} {\frac{{{{\left| {{{\bf{\hat h}}_{jjm}^{H}}{\bf{\hat h}}_{jjk}} \right|}^2}}}{M}},\\
\label{Eq15}&{P_{i,out}}=\frac{1}{{KL_{p}}}\sum\limits_{l \in {{\mathbb{S}}}_j^{p,d}} {\sum\limits_{k = 1}^K {\frac{{{{\left| {{\bf{\hat h}}_{ljm}^H{{{\bf{\hat h}}}_{llk}}} \right|}^2}}}{{{M^2}}}} },\\
&{P_{e}} = \frac{1}{{K(L_{p}+1)}}\sum_{l \in \{j\}\cup{{\mathbb{S}}}_j^{p,d}}\sum_{k=1}^{K}{\rm E}\left\{\frac{{\bf \hat h}_{llk}^{H}{{\bf{\tilde \Theta }}_{ljm} }{\bf \hat h}_{llk}}{M}\right\} \label{Eq16}, \end{align}}which are respectively the normalized power of the desired signal, intra-cell interference, inter-cell interference, and effective noise created by unknown CSI error. Their different normalization factors are used to guarantee that the mean of each term is bounded as shown in the next section. In (\ref{Eq16}), the expectation is over the channel estimations ${\bf \hat h}_{llk}$ and the channel estimation error ${\bf \tilde h}_{llk}^{H}$ since they are unknown at the receiver. 
In the derivations of \eqref{Eq15} and \eqref{Eq16}, we utilize ${{{\bf{h}}_{ljm}^{H}}{\bf{\hat h}}_{llk}} = 0$ for $l \in {{\mathbb{S}}}_j^{o,d}$
based on \eqref{Eq4}, \eqref{Eq5} and \eqref{Eq9}.

With the results in \eqref{Eq13}-\eqref{Eq16}, the SINR of User $m $ in Cell $j$ can be written as:
\begin{equation}\label{Eq17}
\begin{array}{l}
\hspace{-12pt}{\rm SINR}_{jm} = M\frac{{{P_{s}}}}{{\left( {K - 1} \right){P_{i,in}} +M{KL_{p} }{P_{i,out}} + {K(L_{p}+1)P_e} + \frac{KQ}{{\rho}}}}
\end{array}.
\end{equation}
The achievable ergodic rate is thus
\begin{equation}\label{Eq18}
  {C_{jm}} = {\rm E}\left\{ {\frac{}{}{{\log }_2}\left( {1 + {\rm SINR}_{jm}} \right)} \right\}.
\end{equation}

\section{Mean Square Convergence based Achievable Rate Scaling Law}
This section studies the general performance scaling law of the massive MIMO network. First, the means and SCVs of all random SINR components are calculated. Consequently, a sum-rate lower bound based on Jensen's inequality is derived via exploiting the asymptotically deterministic property of the desired signal power. Then, the general performance scaling law is obtained and typical network scenarios for the SINR to be non-decreasing are discussed. Finally, conditions for the scaling law to be applicable for large but finite antenna numbers and extended discussions on the possible effect of PCE are both given.

\subsection{Sum-Rate Lower Bound and Asymptotically Effective SINR}
Different from directly using the almost sure convergence based lemmas \cite{Label9}, \cite{Label1}, \cite{Label2} to calculate the deterministic equivalence of \eqref{Eq17}, we consider the mean square convergence \cite{Label33} and use the SCV (defined as the square of the ratio of the standard deviation over the mean of a random variable) in the analysis \cite{Label19,Label33}. With the use of SCV and mean square convergence, the convergence analysis of the complicated SINR can be transferred to that of its components, which is more tractable. Moreover, it can help us to quantize the convergence speed of the SINR and its components. Specifically, for a random variable sequence $X_M$ with a bounded mean, it converges in mean square to a deterministic value $x$, denoted as, ${X_M}\mathop  \to \limits^{m.s.} x$, if $\mathop {\lim }\limits_{M \to \infty } {\rm Var}\left\{ {{X_M}} \right\} = 0$. Further, $X_M$ is said to be asymptotically deterministic in the sense of mean square convergence if $\mathop {\lim }\limits_{M \to \infty } {\rm SCV}\left\{ {{X_M}} \right\} = 0$ \cite{Label33}.

To utilize the mean square convergence, we first calculate the means and SCVs of all random SINR components and simplify $P_e$. The following lemma is obtained.
\begin{lemma}\label{Lemma1}
\begin{align}
\label{Eq19}&{\rm E}\{P_{s}\}=\mathcal{O}(Q^2),\quad{\rm SCV}\{P_{s}\}=\mathcal{O}\left(\frac{4}{Mc}\right);\\
\label{Eq20}&{\rm E}\{P_{i,in}\}=\frac{{{Q^2}}}{c},\quad \hspace{-2mm}{\rm SCV}\{P_{i,in}\}=\mathcal{O}\hspace{-1mm}\left(\frac{1}{Mc } + \frac{1}{{K - 1}}\right);\\
&{\rm E}\left\{ {{P_{i,out}}} \right\} = {\alpha ^2}Q^2\left( {\frac{1}{K} + \frac{1}{{Mc}}} \right), \nonumber \\
\label{Eq21}&{\rm SCV}\left\{ {{P_{i,out}}} \right\} =\frac{1}{{{L_{p}}Mc}}\frac{{\left( {4 + \frac{{5\left( {K + 1} \right)}}{{Mc}} + \frac{{{K^2} + K + 4}}{{{M^2}{c^2}}}} \right)}}{{{{\left( {1 + \frac{K}{{Mc}}} \right)}^2}}}; \\
&P_{e} = \frac{{{Q}}}{{({L_{p}} + 1)c}}\left( {1 - {Q} + \alpha \left( {1 - \alpha {Q}} \right){L_{p}}} \right). \label{Pe}
\end{align}

\end{lemma}
\begin{IEEEproof}
See Appendix \ref{Appendix A}.
\end{IEEEproof}
\begin{remark}\label{Remark 1}
By noticing that $c \in ( {0,1}]$, $Q, \alpha\in (0, 1)$, and $L_{p}\in[0,L-1]$, it can be seen that the random variables $P_{s}$, $P_{i,in}$ and $P_{i,out}$ all have bounded means. Moreover, apparently the SCV of $P_{s }$ approaches 0 as $M \to \infty $ with a linear convergence rate. Thus for large $M$, $P_{s}$ can be approximated with its mean. However, for $P_{i,in}$ and $P_{i,out}$, their SCVs depend on the scaling of $K$ and $L_{p}$ which are not necessarily $\mathcal{O}(1/M)$. Thus, at this point, we do not use their mean values for further derivations, but keep their random nature for more careful analysis.
\end{remark}

Therefore, the SINR expression in \eqref{Eq17} becomes
\begin{equation}\label{Eq23}
\begin{array}{l}
\hspace{-5pt}{\rm SINR}_{jm} \approx \frac{{{MQ^2}}}{{\left( {K - 1} \right){P_{i,in}} +M{KL_{p} }{P_{i,out}} + {K(L_{p}+1)P_e} + \frac{KQ}{{\rho}}}}
\end{array},
\end{equation}
based on which the following result on the achievable rate can be obtained.
\begin{lemma}\label{Lemma 2}
The achievable rate of User $m$ in Cell $j$ in the massive MIMO network has the following lower bound:
\begin{equation}\label{Eq24}
{C_{jm}} \ge {C_{jm,LB}} \buildrel \Delta \over = {\log _2}\left( {1 + {\rm {\widetilde {SINR}}}_{jm}} \right),
\end{equation}
where
\begin{equation}\label{Eq25}
{\rm {\widetilde {SINR}}}_{jm}= \frac{1}{{\frac{K}{{M{Q}c}}\left( {1 + \alpha {L_{p}}} \right) + {L_{p}}{\alpha ^2} - \frac{1}{{Mc}} + \frac{K}{{M{Q}\rho }}}}.
\end{equation}
\end{lemma}
\begin{IEEEproof}
As ${\log _2}\left( {1 + {1 \mathord{\left/
 {\vphantom {1 x}} \right.
 \kern-\nulldelimiterspace} x}} \right)$ is a convex function of $x$ \cite{Label22}, according to Jensen's inequality, we have
\begin{equation}\label{Eq26}
{C_{jm}} \ge {\log _2}\left( 1 + {{\rm E}\left\{{{\rm SINR}_{jm}^{-1}}\right\}}^{-1} \right).
\end{equation}
By applying the SINR approximation in \eqref{Eq23}, it can be shown that
\begin{eqnarray}
&&\hspace{-10mm}{{\rm E}\left\{{{\rm SINR}_{jm}^{-1}}\right\}}^{-1} = \nonumber\\
&&\hspace{-7mm}\frac{{MQ^2}}{{\rm E}{\left\{ \hspace{-1mm}{\left(K \hspace{-0.5mm}- \hspace{-0.5mm}1 \right)\hspace{-0.5mm}P_{i,in}\hspace{-0.5mm} + \hspace{-0.5mm}M\hspace{-0.5mm}K\hspace{-0.5mm}{L_{p}}\hspace{-0.5mm}{P_{i,out}} \hspace{-0.5mm}+ \hspace{-0.5mm}K\hspace{-0.5mm}({L_{p}}\hspace{-0.5mm} +\hspace{-0.5mm} 1)\hspace{-0.5mm}{P_e}\hspace{-0.5mm} +\hspace{-0.5mm} \frac{{K{Q}}}{\rho }} \hspace{-0.5mm}\right\}}}. \label{Eq27}
\end{eqnarray}
By defining the effective asymptotic SINR as ${\rm {\widetilde {SINR}}}_{jm}={{\rm E}\left\{{{\rm SINR}_{jm}^{-1}}\right\}}^{-1}$ and directly using the results in \eqref{Eq20}-\eqref{Pe}, the lower bound in \eqref{Eq24} can be obtained.
\end{IEEEproof}

In existing works \cite{Label28,Label29}, closed-form approximations of the ergodic rate were derived by using \cite[Lemma 1]{Label3}, i.e., ${\rm E}\{\log_2(1+\frac{X}{Y})\}\approx \log_2(1+\frac{{\rm E}\{X\}}{{\rm E}\{Y\}})$, based on the law of large numbers \cite[Lemma 1]{Label3}. In this work, we utilize the asymptotically deterministic property of the desired signal power term in the SINR expression and Jensen's inequality to derive a lower bound of the ergodic rate, which has the same format as the SINR approximations in \cite{Label28,Label29}.

Due to \eqref{Eq24}, understanding the scaling law of the achievable sum-rate  is transformed to understanding the scaling law of
${\rm {\widetilde {SINR}}}_{}$ where the subscript of user index and cell index in \eqref{Eq25} is omitted due to homogeneous network assumptions.
\subsection{Scaling-Law Results}\label{scalinglawres}
In this subsection, the scaling law of the asymptotically effective SINR is analysed to
show how the system performance is affected by network parameters and pilot contamination level. For all system parameters, including the number of users $K$, the transmission power $\rho$, and the channel training energy $E_t$, a general scaling  model with respect to the BS antenna number $M$ is used.
Note that other system parameters, i.e., the channel spatial correlation metric $c$, the normalized  inter-cell large scale fading $\alpha$ are constant with respect to $M$.

Assume that
\begin{eqnarray}
&&K =\mathcal{O}\left( {{M^{{r_k}}}} \right),
\quad \frac{1}{\rho} =\mathcal{O}\left( {{M^{{r_\rho}}}} \right), \nonumber \\
&&\frac{1}{{{E_t}}} = \mathcal{O}\left( {{M^{{r_t}}}} \right),
\quad{\rm {\widetilde {SINR}}} = \mathcal{O}\left( {{M^{{r_s}}}} \right),
\label{Eq28}
\end{eqnarray}
where the exponents $r_k$, $r_\rho$, and $r_t$ represent the scales of $K$, $1/{\rho}$, and $1/{E_t}$ with respect to $M$; and $r_s$ represents the scale of the effective SINR. For practical ranges of the network parameters, it is assumed that $0 \le {r_k},{r_\rho},{r_t}\le 1$. The reasons are given as follows.

Firstly, in typical applications of massive MIMO, $K$ either increases with $M$ or keeps constant. Thus ${r_k} \ge 0$. On the other hand, $K$ cannot exceed $M$ since the maximum multiplexing gain  is $M$. Thus, ${r_k} \le 1$. Secondly, due to the high energy efficiency requirement of massive MIMO\cite{Label2,Label4}, the transmission power $\rho$ should not increase with $M$. But it can decrease as $M$ increases with the condition that its decreasing rate is lower than the increasing rate of $M$. This is because that the maximum achievable array gain for compensating the loss of receiving energy is $M$. Thus $0 \le {r_\rho} \le 1$. For the same reasons, $0 \le {r_t} \le 1$ is considered.

Another important parameter for the performance analysis  is $L_{p}$ which quantifies the pilot contamination level. Existing works \cite{Label9}, \cite{Label1} show that pilot contamination does not vanish as $M$ grows for i.i.d. channel \cite{Label9} or spatially correlated channel with uniform channel directions \cite{Label1}.
For more practical channel models, by exploiting the large spatial degrees of freedom provided by the massive antenna array at the BS, several efficient methods have been proposed to reduce pilot contamination \cite{Label12,Label34,Label35,Label13}. Specifically, the difference among users' channel directions \cite{Label12} or locations \cite{Label34} and smart pilot allocation based on them \cite{Label35} were utilized to avoid pilot contamination, and corresponding simulations showed that as $M$ increases, lower pilot contamination level can be achieved. In our work, the effect of PCE is reflected by the $L_p$ parameter in our modeling, which is assumed to be a constant. The special case of $L_p=0$ represents perfect PCE\footnote{This case can happen for systems with small $c$, i.e., highly correlated channels.}, while $L_p\ne 0$ represents that pilot contamination is not fully annihilated and on average there are $L_p$ cells with pilot contamination to the cell of interest.

With the above scaling model, the general performance scaling law is given as follows.

\begin{theorem}\label{Theorem 1}
For the massive MIMO network with MMSE channel estimation and MRT, if perfect PCE is achieved, the performance scaling law is
\begin{equation}\label{Eq30}
{r_s} = 1 - {r_t} - {r_k} - {r_\rho}.
\end{equation}
For imperfect PCE, the performance scaling law is
\begin{equation}\label{Eq31}
{r_s}= \min \left( {1 - {r_t} - {r_k} - {r_\rho},0} \right).
\end{equation}

\end{theorem}
\begin{IEEEproof}
The scaling exponent of ${\rm {\widetilde {SINR}}}$ in \eqref{Eq25} is determined by the maximal scaling exponent of the terms in its denominator. Directly from \eqref{Eq25}, it can be seen that given aforementioned parameter range, either $L_{p}{\alpha ^2}$ or $K/(MQ\rho)$ or both have the highest order with respect to $M$. Further, since $Q$ and $E_t$ have the same scaling from \eqref{Eq8}, (\ref{Eq30}) and (\ref{Eq31}) can be obtained from (\ref{Eq28}).
\end{IEEEproof}

Since having decreasing performance  with respect to $M$ contradicts the motivations of massive MIMO, we present the condition for \textit{non-decreasing} SINR (i.e., ${r_s} \ge 0$ ) in the following corollary.
\begin{corollary}\label{Corollary 1}
For both perfect and imperfect PCE, the necessary and sufficient condition for the massive MIMO network with MMSE channel estimation and MRT to have non-decreasing SINR is
\begin{equation}\label{Eq32}
{r_t} + {r_k} + {r_\rho} \le 1,\hspace{0.3cm}{r_t},{r_k},{r_\rho} \in \left[ {0,1} \right]
\end{equation}
\end{corollary}
\begin{IEEEproof}
The result is a straightforward extension from Theorem \ref{Theorem 1}.
\end{IEEEproof}

The scaling law in (\ref{Eq30}) and (\ref{Eq31}) illustrates quantitatively the relationship between the performance scaling and scalings of all important parameters. The condition in (\ref{Eq32}), on the other hand, provides useful guidelines for the design of the massive MIMO network.

\begin{remark}
Equation (\ref{Eq30}) shows that with perfect PCE, the scaling of the SINR $r_s$ is determined by the scaling exponent of the channel training energy $r_t$ and per-user transmission power $r_k+r_\rho$. Moreover, (\ref{Eq30}) also shows that the scaling of the SINR is a decreasing function of both. Thus higher per-user transmission power and training energy result in improved performance, and one can compensate for the other in performance. (\ref{Eq31}) shows that with imperfect PCE, $r_s$ is upper bounded by zero due to the pilot contamination bottleneck and it can be seen from \eqref{Eq25} that ${\rm {\widetilde {SINR}}}\le 1/(L_p\alpha^2)$. This means that the asymptotic performance will be constant with respect to $M$, while the concrete value is dependent on the efficiency of PCE, which is more general than the result in \cite{Label1}.
\end{remark}

\subsection{Discussions on Typical Massive MIMO Settings}\label{Dis_sec_1}
In this subsection, the scaling law in (\ref{Eq30}) and (\ref{Eq31}) and the condition for non-decreasing SINR in (\ref{Eq32}) are elaborated for typical network settings.
\begin{itemize}
\item [D1]\label{D1}
First, we consider the case of $r_t=0$, i.e., the training energy $E_t$ is constant. For perfect PCE, $r_s=1-(r_k+r_\rho)$. For imperfect PCE, $r_{s}=\min(1-(r_k+r_\rho), 0)$. To have non-decreasing SINR, $r_k+r_\rho\le1$ is needed. This shows a tradeoff between the scaling of the user number $K$ and the transmission power $\rho$. The most power-saving design is to make the per-user transmission power decrease linearly with $M$, i.e., $r_k+r_\rho=1$. With a larger $M$, the network can serve more users or have less power consumption, while maintaining certain performance. However, improvements in the two aspects both have a limit: 1) $r_k=1$, $r_\rho=0$ and 2) $r_k=0$, $r_\rho=1$. Case 1) means that when $K$ increases linearly with $M$, to achieve non-decreasing SINR, $\rho$ must remain constant, and thus the goal of reducing $\rho$ cannot be achieved. Case 2) means that when $\rho$ is scaled inversely proportional to $M$, the goal of serving more users cannot be achieved. The latter case is the major power scaling
scenario considered in the literature \cite{Label1,Label2}. Obviously, our results cover this case, and shows more insights for general scalings of $K$ and $\rho$.
\item [D2]\label{D2}  Then we consider the case of $r_t=1$, i.e., $E_t=\mathcal{O}(1/M)$. For perfect PCE, $r_{s}=-(r_k+r_\rho)$. For imperfect PCE, $r_{s}=\min(-(r_k+r_\rho), 0)$. To have non-decreasing SINR, $r_\rho = r_k = 0$ is needed, i.e., the transmission power and the user number should both remain constant. This shows that the training energy is key to the performance, i.e., with low training energy, the promising features of  massive MIMO
cannot be achieved.
\item [D3]\label{D3}  For the general case where ${r_t} \in \left( {0,1} \right)$, non-decreasing SINR requires ${r_k} + {r_\rho} \le 1 - {r_t}$. That is, the scale of the per-user transmission power should be no less than $1/M^{1-r_t}$. This shows the trade-off between the training phase and the data transmission phase.
\item [D4]\label{D5}  In previous discussions, the scaling of the training energy $r_t$ is treated as a free parameter, next, the  special but commonly used training case where $P_{t} = \rho/K$ and $\tau= K$ is considered, i.e., each user uses the same power in training as that for data transmission. Since in this case, $r_t=r_{\rho}$, we have $r_s=1-2r_{\rho}-r_k$ for perfect PCE and $r_s=\min(1-2r_{\rho}-r_k,0)$ for imperfect PCE. Non-decreasing SINR requires $2r_\rho+r_k\le1$. If further $K$ is constant, $r_\rho\le1/2$ is needed, i.e., the transmission power should scale as $1/\sqrt{M}$ or higher. This is the same as the results in \cite{Label1}, \cite{Label2}.
\item [D5]\label{D6}  Another typical setting from the perspective of supporting a large number of users is to have $K$ increase linearly with $M$, i.e., $r_k = 1$. For this case,  $r_s=-r_t-r_\rho$ for perfect PCE and $r_s=\min(-r_t-r_\rho,0)$ for imperfect PCE. Non-decreasing SINR requires $r_t = r_\rho = 0$. Thus, to support such a number of users, the training energy must be constant and the transmission power cannot decrease with $M$ simultaneously.
\end{itemize}

\subsection{Applicability of Scaling Law for Large but Finite $M$}
The performance scaling law is derived for $M\rightarrow\infty$. To be practically useful, its effectiveness for large but finite antenna numbers needs to be understood. In our scaling law derivation, the term with the highest scaling exponent in the denominator of \eqref{Eq25} is kept while other lower order terms are neglected. Thus for the interested range of antenna number, the result is said to be applicable if the term with the highest order is dominant, in other words, significantly larger (denoted by the symbol ``$\gg$'') than the rest\footnote{Although there is no quantitative definition on dominance, generally speaking, a positive number $b$ is said to be dominated by another positive number $a$, i.e., $a\gg b$, when $a$ is at least one-order high than $b$, or in other words, $a\ge 10b$. This common practice is used in our work.}. In the following, we derive conditions for the scaling law to be applicable in practical systems with large but finite antenna numbers. To our best knowledge, there has been no analytical studies on the applicability of asymptotic scaling law in existing work.
\begin{corollary}\label{corollary 2}
For networks with perfect PCE, the scaling law in (\ref{Eq30}) is applicable for large but finite $M$ when
\[\frac{1}{\rho}\gg\frac{1}{c}(1-\frac{Q}{K}).\]
\end{corollary}
\begin{IEEEproof}
In this case, $L_p=0$ and the term with the highest order with respect to $M$ is $K/(M{Q}\rho )$. For it to be dominant in the denominate of (\ref{Eq25}), it can be seen straightforwardly that $\frac{K}{Q\rho}\gg\frac{1}{c}(\frac{K}{Q}-1)$ is sufficient.
\end{IEEEproof}
\begin{corollary}\label{corollary 3}
For networks with imperfect PCE, the scaling law in (\ref{Eq31}) is applicable for large but finite $M$ when the following holds:
\begin{equation}
\left\{\begin{array}{ll}
\hspace{-2mm}\frac{K}{MQ\rho } \gg \frac{1}{Mc}\chi  + {L_p}{\alpha ^2} &\hspace{-1mm} \text{if } 1 - {r_t} - {r_k} - {r_\rho } < 0 \\
\hspace{-2mm} L_p\alpha ^2 \gg \frac{1}{M}(\frac{1}{c}\chi  \hspace{-1mm} +\hspace{-1mm}\frac{K}{Q\rho }) & \hspace{-1mm}\text{if } 1 - {r_t} - {r_k} - r_\rho  > 0\\
\hspace{-2mm}{L_p}{\alpha ^2} \gg \frac{1}{M}(\frac{1}{c}\chi \hspace{-1mm} +\hspace{-1mm} \frac{K}{Q\rho })\text{or} \frac{K}{Q\rho } \gg \frac{1}{c}\chi & \hspace{-1mm}\text{if } 1 - {r_t} - {r_k} - {r_\rho } = 0.
\end{array}\right.
\end{equation}
where $\chi=\frac{K}{Q}(1+\alpha L_p)-1$.
\end{corollary}
\begin{IEEEproof}
The derivation is almost the same as the one for Corollary \ref{corollary 2}. Thus it is omitted.
\end{IEEEproof}
\begin{remark}\label{Remark 5}
Corollary \ref{corollary 2} shows that for large but finite $M$, the scaling law in (\ref{Eq30}) for perfect PCE is more accurate for less correlated channel (larger $c$) and/or lower transmission power. This is because the effective antenna number increases as $c$ increases. On the other hand, when $\rho$ decreases, the interference power decreases while the effect of noise power becomes more dominant. This transmission power sensitivity can be also found in the simulation results on the power scaling law in \cite{Label2}. Corollary \ref{corollary 3} shows that for large but finite $M$, the scaling law in (\ref{Eq31}) for imperfect PCE is also more accurate for less correlated channel, while the effect of the transmission power on the scaling law accuracy is unclear.

\end{remark}

The sufficient conditions\footnote{For example, for perfect PCE, the denominate of \eqref{Eq25} can be written as $\frac{K}{{MQ}}\left( {\frac{1}{c}{\rm{ + }}\frac{1}{\rho }} \right) - \frac{1}{{Mc}}$. Thus, another case for the scaling law in (\ref{Eq30}) to be applicable is $r_\rho=0$ and $\frac{K}{Q} \gg \frac{1}{1+\frac{c}{\rho}}$.} in Corollaries \ref{corollary 2} and \ref{corollary 3} for the scaling laws to be applicable provide guidance for using the results \eqref{Eq30} and \eqref{Eq31} in practical system design with large but finite antenna number. For systems that satisfy the conditions in Corollaries \ref{corollary 2} and \ref{corollary 3}, the derived scaling law can accurately reflect the network performance; while when the conditions are not guaranteed, the result becomes less accurate but can still provide rough suggestions on the performance. Consider the following two system settings 1) $\rho=1/\sqrt{M}$ and 2) $\rho=20/\sqrt{M}$, where the other system parameters are as follows: $L_p=0$, $c=0.6$, $E_t=10$ dB, $K=10$. When $M$ is in the range of $[200,600]$, the ratio of $\frac{1}{\rho}$ to $\frac{1}{c}(1-\frac{Q}{K})$ approximately belongs to the interval [9.4, 16.2] for Setting 1 and the interval [0.47, 0.81] for Setting 2. Thus the scaling law is applicable to Setting 1 but not to Setting 2. This will be shown by the simulation results in Fig. 1 of Section \ref{simulation}.


\subsection{Effect of Decreasing Pilot Contamination due to PCE}\label{decreasingPC}
In the scaling law analysis in Section \ref{scalinglawres}, $L_p$, the average number of cells with pilot contamination  is assumed to be a constant. As also mentioned in that section, a decreasing $L_{p}$ with respect to $M$ is possible by PCE. To take into consideration the possible variation of $L_p$ with respect to $M$, the following approximate scaling model for $L_p$ can be used:
\begin{equation}\label{Eq29}
\frac{1}{{{L_{p}}}} \approx \mathcal{O}\left( {{M^{{r_\gamma}}}} \right)
\end{equation}
where $0 \le r_\gamma\le 1$. The value of the exponent $r_\gamma$ is dependent on the concrete algorithm. For example, $r_\gamma=0$ corresponds to the imperfect PCE case discussed previously, where no PCE is available or the pilot contamination is only reduced to a constant value. Since the highest achievable scaling of pilot contamination cannot surpass the increasing rate of antenna number, we have $r_\gamma \le 1$.

With this more general modeling, the corresponding performance scaling law by following similar derivation procedure can be obtained as
\begin{equation}\label{Eq32_170210}
{r_s} \approx \min \left( {1 - {r_t} - {r_k} - {r_\rho},r_\gamma} \right).
\end{equation}

\begin{remark}\label{Remark 6_170211}
The expressions in \eqref{Eq29} and \eqref{Eq32_170210} are approximations since $L_{p}$ is modeled as an integer in this paper. Its variation with respect to $M$ can be roughly fitted with the function $a/M^{r_\gamma}$ for a constant $a$.
When $L_{p}$ decreases with $M$ (meaning high PCE efficiency and a positive $r_\gamma$), an increasing effective SINR rather than the bounded one in \eqref{Eq31} can be obtained if $1 - {r_t} - {r_k} - {r_\rho}>0$, which will be evaluated by simulations in Section \ref{simulation}.

\end{remark}

\section{Conditions for the SINR to be Asymptotically Deterministic}
One important concept in massive MIMO analysis is asymptotically deterministic. For example, when the desired signal power and interference power which are random in finite-dimension converge to deterministic values as $M$ increases to infinity \cite{Label9,Label10}, their expected values can be used to replace the random counterparts to simplify the analysis. Then a natural question is when the massive MIMO system has asymptotically deterministic SINR for the corresponding performance analysis to be reliable. In existing literature, the deterministic equivalence is mainly based on the almost sure convergence\cite{Label1}. While many results have been reported based on the aforementioned analysis, the corresponding convergence condition tends to be vague and explicit analysis of the effect of parameter scalings on the convergence is missing.

In this section, following the work in \cite{Label33}, we adopt the mean square convergence \cite{Label18} for the definition of asymptotically deterministic property, and use the SCV scaling with respect to  $M$ to derive sufficient conditions for asymptotically deterministic SINR. Finally, typical network scenarios are discussed.
\begin{definition}\label{Definition 2}
The random variable sequence $X_M$ is said to be asymptotically deterministic if its SCV decreases linearly with $M$ or faster.
\end{definition}

Strictly speaking, there is no constraint on the scaling of the SCV  for a random sequence to converge to a constant value in the mean square sense other than that the SCV converges to 0 as $M \to \infty$. Notice that for any positive number $\alpha$, ${1 \mathord{\left/
 {\vphantom {1 {{M^\alpha } \to 0}}} \right.
 \kern-\nulldelimiterspace} {{M^\alpha } \to 0  }}$ when $M \to \infty$. However, the convergence can be very  slow for small $\alpha$, in which case, the derived asymptotic results may not be applicable for practical range of large but finite $M$. Thus, for practical system applications, the SINR is only considered to be  asymptotically deterministic when the SCV decreases linearly with $M$ or faster.
With this definition, we actually derive sufficient conditions for the SINR to be asymptotically deterministic.
\begin{proposition}\label{Pro1}When $M \gg 1$, for perfect PCE, a sufficient condition for the SINR in \eqref{Eq17} to be asymptotically deterministic is
\begin{equation}\label{Eq33}
2r_t+r_k+2r_\rho \ge 1.
\end{equation}
For imperfect PCE,
the SINR is always asymptotically deterministic.
\end{proposition}

\begin{IEEEproof}
Please see Appendix \ref{Appendix B}.
\end{IEEEproof}

Note that the above sufficient conditions are based on the scaling law in \eqref{Eq30} and \eqref{Eq31}. They can be utilized to justify whether the cases discussed in D1-D5 (including the power scaling analysis in existing works \cite{Label1,Label2,Label3}) satisfy the condition for asymptotically deterministic SINR. For perfect PCE, \eqref{Eq33} implies $r_s\le1/2$, meaning that to make the SINR asymptotically deterministic, the highest possible SINR scaling is $\sqrt{M}$.
Without the constraint for the SINR to be asymptotically deterministic, the maximum possible value for $r_s$ is $1$. For imperfect PCE, Proposition \ref{Pro1} implies that no new limit is casted to
the SINR scaling for the SINR to be asymptotically deterministic.


\begin{remark}
The asymptotically deterministic definition of this work is based on mean square convergence with extra convergence speed requirement that the SCV decreases linearly with $M$ or faster. This is different to the analysis in \cite{Label9,Label1}, where almost sure convergence was used via the law of large numbers without specification on the convergence speed. In  \cite{Label9}, it was shown that for the case of $L_p=L-1$ the SINR converges to $1/(L_p\alpha^2)$ when  $M\rightarrow\infty$ with a fixed $K$. First, Proposition \ref{Pro1} of this work shows that the SINR is asymptotically deterministic under Definition \ref{Definition 2}. Second,  for constant $K$ case or even further when the user number has ``faster than linear" decay ($r_k<1$),  it can be shown easily that (\ref{Eq25}) converges to $1/(L_p\alpha^2)$ when $M\rightarrow \infty$. In \cite{Label1}, several linear precoders were studied for $M,K\rightarrow\infty$ with fixed $M/K$, corresponding to $r_k=1$. It was shown that every term of the SINR expression is asymptotically deterministic based on which the deterministic equivalence of the SINR and the asymptotic sum-rate were derived. Proposition \ref{Pro1} of this work shows that the condition  $r_k=1$ guarantees the SINR to be asymptotically deterministic under Definition \ref{Definition 2}. And the derived asymptotic SINR in \cite{Label1} has the similar format to that in (\ref{Eq25}).
\end{remark}

Next, typical scenarios with asymptotically deterministic SINR are investigated in which $r_t$ is allowed to take values from $\{0,1/2,1\}$ only. The tradeoff between parameters will be revealed.
\begin{itemize}
\item [D6]\label{D7}  For perfect PCE, to achieve both $r_s = 1/2$ (i.e., the SINR increasing linearly with $\sqrt{M}$) and asymptotically deterministic SINR, the sufficient condition reduces to $r_k = 0$, ${r_t} + {r_\rho } = 1/2$. It means that the user number should be constant, the product of training energy and transmission power must scale as $1/\sqrt{M}$ where the tradeoff between the energy efficiency in training and transmission periods can be found. Two typical cases with constant user number may happen: a) ${r_t}=0, {r_\rho }=1/2$; and b) ${r_t}=1/2, {r_\rho }=0$. For Case a), when the training energy is constant, the transmission power should scales as $1/\sqrt{M}$. For Case b), when the training energy scales as $1/\sqrt{M}$, the transmission power should be constant.

\item [D7]\label{D8}
For imperfect PCE, to achieve both the highest scaling ${r_s} = 0$ and asymptotically deterministic SINR, the sufficient condition is the same as the condition for non-decreasing SINR in \eqref{Eq32}, i.e., $r_k+r_\rho+r_t\le1$. It means that the product of the training energy and the per-user transmission power should scale as $1/M$ or higher. Typical cases are referred to former discussions in Section \ref{Dis_sec_1}.
\item [D8]\label{D9}
For perfect PCE, to achieve constant and asymptotically deterministic SINR, the sufficient condition reduces to ${r_t} + {r_k} + {r_\rho } = 1$. It means that the product of training energy and the per-user transmission power must scale as $1/M$. Three typical cases may happen: a) ${r_t}=0, {r_k+r_\rho }=1$; b) ${r_t}=1/2, {r_k+r_\rho}=1/2;$ and c) ${r_t}=1, {r_k+r_\rho}=0$. For Case a), when the training energy is constant, the per-user transmission power should scales as $1/M$. For Case b), when the training energy scales as $1/\sqrt{M}$, the per-user transmission power should also scales as $1/\sqrt{M}$. For Case c), when the training energy scales as $1/M$, the per-user transmission power should be constant.
\end{itemize}

\begin{remark}
In existing works, only constant SINR case ($r_s = 0$) has been considered \cite{Label1}. The discussion in D6-D8 shows that for perfect PCE, the asymptotically deterministic SINR can scale as $\sqrt{M}$. For imperfect PCE, at most constant asymptotically deterministic SINR can be achieved without any extra conditions besides the one for non-decreasing SINR in \eqref{Eq32}.
\end{remark}


\section{THE ZERO-FORCING COUNTERPART}
In this section, we will give the scaling law and  applicability results for the ZF precoder.
\subsection{Downlink Transmission with ZF}
With the channel estimation in (\ref{Eq7}), the ZF precoding matrix at BS $l$ is
${\bf W}_{ll}={\bf \hat H}_{ll}\left({\bf \hat H}_{ll}^H{\bf \hat H}_{ll}\right)^{-1}$,
where ${\bf \hat H}_{ll}=[{\bf \hat h}_{ll1},...,{\bf \hat h}_{llK}]$.
The received signals at $K$ users in Cell $j$ can be written as
\begin{eqnarray}
&&\hspace{-1cm}{\bf y}_{j}\hspace{-3pt}=\hspace{-3pt}\sqrt{\rho \lambda}{\bf x}_j+\sqrt{\rho \lambda}{\bf \tilde H}_{jj}^H{\bf W}_{jj}{\bf x}_j \nonumber \\
&&+\sum\limits_{l\ne j}\hspace{-3pt}\sqrt{\rho\lambda}\left({\bf \hat H}_{lj}^{H}+{\bf \tilde H}_{lj}^{H}\right){\bf W}_{ll}{\bf x}_l+{\bf n}_{j},
\label{zf_Eq1}
\end{eqnarray}
where ${\bf \hat H}_{lj}=[{\bf \hat h}_{lj1},...,{\bf \hat h}_{ljK}]$ is the estimation of the channel from users in Cell $j$ to BS $l$, ${\bf \tilde H}_{lj}=[{\bf \tilde h}_{lj1},...,{\bf \tilde h}_{ljK}]$ is the corresponding channel estimation error,  ${\bf n}_{j}\sim \mathcal{C}\mathcal{N}({\bf 0},{\bf I}_K)$ is the noise vector at the user side,  and $\lambda$ is the power constraint coefficient. Since ${\bf \hat H}_{ll}^H{\bf \hat H}_{ll}\cong \frac{Q}{c}{\bf X}_{ll}$ with ${\bf X}_{ll}\sim W_K(\Delta, {\bf I}_{K})$ from \eqref{Eq9}, it can be shown that ${\rm E}\{{\rm tr}\{{\bf X}_{ll}^{-1}\}\}=K/(\Delta-K)$ \cite{Label2} and thus
\[\lambda={{\rm E}\{{\rm tr}\{({\bf \hat H}_{ll}^H{\bf \hat H}_{ll})^{-1}\}\}}^{-1}=\frac{MQ(\Delta-K)}{\Delta K}.\]

Since ${\bf h}_{ljm}^H{\bf \hat h}_{llk}=0, \forall l\in {{{\mathbb{S}}}_j^{o,d}}$ , \eqref{zf_Eq1} becomes
\begin{eqnarray}
&&\hspace{-1cm}{\bf y}_{j}=\sqrt{\rho \lambda}{\bf x}_j+\sqrt{\rho \lambda}{\bf \tilde H}_{jj}^H{\bf W}_{jj}{\bf x}_j\nonumber\\
&&+\sum\limits_{l\in {{{\mathbb{S}}}_j^{p,d}}}\hspace{-7pt}\sqrt{\rho\lambda}({\bf \hat H}_{lj}^{H}+{\bf \tilde H}_{lj}^{H}){\bf W}_{ll}{\bf x}_l+{\bf n}_{j}.
\label{zf_eq36}
\end{eqnarray}
From \eqref{Eq11}, we have ${\bf \hat h}_{ljk}=c\beta_{ljk}{\bf \hat h}_{llk}, \forall j \in \{l\}\cup {{\mathbb{S}}}_l^{p,s}$ and
\begin{equation}
  {\bf \hat H}_{lj}=\alpha {\bf \hat H}_{ll}, \forall j \in {{\mathbb{S}}}_l^{p,s}.
\end{equation}
Thus, \eqref{zf_eq36} can be further simplified to
\begin{eqnarray}
&&\hspace{-1cm}  {\bf y}_{j}=\sqrt{\rho \lambda}{\bf x}_j+\sum\limits_{l\in {{{\mathbb{S}}}_j^{p,d}}}\sqrt{\rho\lambda}\alpha{\bf x}_l\nonumber \\
&&+\sum\limits_{l\in {\{j\} \cup {{\mathbb{S}}}_j^{p,d}}}\hspace{-10pt}\sqrt{\rho\lambda}{\bf \tilde H}_{lj}^{H}{\bf W}_{ll}{\bf x}_l+{\bf n}_{j}
\end{eqnarray}
and
\begin{eqnarray}\label{zf_eq39}
&&\hspace{-1cm}
 {y}_{jm}=\sqrt{\rho \lambda}{x}_{jm}+\sum\limits_{l\in {{{\mathbb{S}}}_j^{p,d}}}\sqrt{\rho\lambda}\alpha{x}_{lm}\nonumber \\
 &&+\sum\limits_{l\in {\{j\} \cup {{\mathbb{S}}}_j^{p,d}}}\hspace{-10pt}\sqrt{\rho\lambda}{\bf \tilde h}_{ljm}^{H}{\bf W}_{ll}{\bf x}_l+{n}_{jm}.
\end{eqnarray}

Similar to the MRT precoding case, the SINR can be expressed as
\begin{equation}
{\rm SINR}_{jm}^{ZF}=\frac{\rho\lambda}{1+\alpha^2\rho\lambda L_p+{\bar P}_e},
\end{equation}where the CSI error term ${\bar P}_e$ can be calculated by following the definition of ${\bf \tilde \Theta}_{ljm}$ in \eqref{Eq10} as follows.
\begin{eqnarray}
{\bar P}_e \hspace{-9pt}&=& \hspace{-7pt} \rho\lambda \hspace{-5pt}\sum\limits_{l\in {\{j\} \cup {{\mathbb{S}}}_j^{p,d}}}\hspace{-5pt}{\rm E}\left\{{\rm tr}\left\{{\bf W}_{ll}^H{\bf \tilde \Theta}_{ljm}{\bf W}_{ll}\right\}\right\} \nonumber\\
\hspace{-9pt}
&=& \hspace{-8pt} \frac{\rho \lambda}{Q}(1-Q){\rm E}\left\{{\rm tr}\left\{{\bf X}_{jj}^{-1}\right\}\right\}\nonumber \\
&&\hspace{1cm}+\frac{\rho \lambda}{Q}\alpha(1-\alpha Q)\hspace{-5pt}\sum_{l\in {{{\mathbb{S}}}_j^{p,d}}}{\rm E}\left\{{\rm tr}\left\{{\bf X}_{ll}^{-1}\right\}\right\} \nonumber \\
\hspace{-9pt}&=& \hspace{-8pt}\frac{\rho}{c}\left[1-Q+\alpha(1-\alpha Q)L_p\right].
\label{zc_pe}
\end{eqnarray}
Further, the SINR can be written as
\begin{equation}\label{zf_effective_SINR}
\hspace{-10pt}{{\rm SINR}^{ZF}}=\frac{1}{\frac{cK}{\rho Q(Mc-K)}+\alpha^2L_p+\frac{K(1+\alpha L_p-Q(1+\alpha^2L_p))}{Q(Mc-K)}},
\end{equation}
which shows that the SINR of ZF is always deterministic due to the elimination of the random interference terms as opposed to that of MRT in \eqref{Eq17}.

\subsection{Scaling-Law Result and Its Applicability}
Similar to the proof of Theorem \ref{Theorem 1}, the scaling exponent of ${\rm {{SINR}}}^{ZF}$ \hspace{-5pt}is determined by the maximal scaling exponent of the terms in its denominator. It can be seen that given aforementioned parameter range, either ${\alpha ^2}L_{p}$, or $\frac{cK}{\rho Q(Mc-K)}$, or both have the highest order with respect to $M$. Therefore, the scaling laws of the ZF precoder for perfect PCE, or imperfect constant PCE, or decreasing PCE are the same as those of the MRT precoder. Related discussions on typical scenarios can be referred to Section \ref{Dis_sec_1}.

\subsubsection{Applicability of Scaling Law for Large but Finite $M$}
Although the MRT precoder and the ZF precoder have the same performance scaling law, their applicability for large but finite $M$ can be different due to the difference in the SINR expressions.
\begin{corollary}\label{corollary zf4}
For networks with perfect PCE, the scaling law in (\ref{Eq30}) is applicable for large but finite $M$ when
the following holds:
\begin{equation}\label{zc_eq44}
\left\{ {\begin{array}{ll}
\frac{1}{\rho}\gg\frac{1-Q}{c} & \text{ if } {r_k} = 1\\
\frac{1}{\rho}\gg\frac{1-Q}{c} \text{ and } Mc\gg K & \text{ if } {r_k} < 1
\end{array}} \right..
\end{equation}
\end{corollary}
\begin{IEEEproof}
With perfect PCE, $L_p=0$. The term with the highest order with respect to $M$ is $\frac{cK}{\rho Q(Mc-K)}$. When $r_k=1$, for it to be dominant in the denominate of \eqref{zf_effective_SINR}, it can be seen straightforwardly that $\frac{1}{\rho}\gg\frac{1-Q}{c}$ is sufficient. Moreover, for the scaling law with $r_k<1$ to be accurate, $\frac{cK}{\rho Q(Mc-K)}\approx\frac{cK}{\rho Q(Mc)}$ i.e., $Mc\gg K$ is further needed.
\end{IEEEproof}
\begin{corollary}\label{zf_corollary5}
For networks with imperfect PCE, the scaling law in (\ref{Eq31}) is applicable for large but finite $M$ when the following holds:
\begin{equation}
\hspace{-2.5mm}\left\{ {\begin{array}{*{20}{c}}
\hspace{-1.2cm}\frac{1}{\rho} \gg \frac{\alpha^2 L_p Q M}{K}+\frac{\tilde{\chi}}{c} \ \text{and} \ Mc\gg K \
&\hspace{-5mm}{\rm{if  }}\hspace{1mm}{1 \hspace{-1mm}-\hspace{-1mm} {r_t} \hspace{-1mm}-\hspace{-1mm} {r_k} \hspace{-1mm}-\hspace{-1mm} {r_\rho } \hspace{-1mm}<\hspace{-1mm} 0}\\
\hspace{-36mm}{L_p}{\alpha ^2} \gg \frac{K\left(\frac{c}{\rho}+\tilde{\chi}\right)}{Q(Mc-K)}
&\hspace{-5mm}{\rm{if  }}\hspace{1mm} {1 \hspace{-1mm}-\hspace{-1mm} {r_t} \hspace{-1mm}-\hspace{-1mm} {r_k} \hspace{-1mm}-\hspace{-1mm} {r_\rho }  \hspace{-1mm}>\hspace{-1mm} 0}\\
\hspace{-2.9mm}{L_p}{\alpha ^2} \hspace{-1mm}\gg\hspace{-1mm} \frac{K\left(\frac{c}{\rho}+\tilde{\chi}\right)}{Q(Mc-K)}\,{\rm{ or }}\,\hspace{-1mm}\left\{\frac{1}{\rho}\hspace{-1mm} \gg \hspace{-1mm}\frac{\tilde{\chi}}{c} \ \text{and} \ Mc\hspace{-1mm}\gg\hspace{-1mm} K \right\}
&\hspace{-3mm}{\rm{if  }}\hspace{1mm}1 \hspace{-1mm}-\hspace{-1mm} {r_t} \hspace{-1mm}-\hspace{-1mm} {r_k} \hspace{-1mm}-\hspace{-1mm} {r_\rho }\hspace{-1mm} =\hspace{-1mm} 0.
\end{array}} \right.
\end{equation}
where $\tilde{\chi}=1+\alpha L_p-Q(1+\alpha^2 L_p)$.
\end{corollary}
\begin{IEEEproof}
The derivation is almost the same as the one for Corollary \ref{corollary zf4}. Thus it is omitted.
\end{IEEEproof}
\begin{remark}\label{zc_remark7}
It can be seen from \eqref{zc_eq44} that the effect of $c$ and $\rho$ on the accuracy of the scaling law for perfect PCE is similar to that of the MRT. One one hand, since the condition in Corollary \ref{corollary 2} is equal to $\frac{1}{\rho}\gg \frac{1-Q/K}{c}$, the low SNR constraint for ZF can be relaxed to some extent. On the other hand, an additional condition $Mc\gg K$ is needed for $r_k<1$ due to the cost of degrees of freedom for interference cancellation in ZF. For imperfect PCE, the scaling law is also more accurate for larger $c$ (note that $(\frac{c}{\rho}+\tilde{\chi})/(Mc-K)$ is a decreasing function of $c$ from $\tilde{\chi}>0$), while the effect of the transmission power on the scaling law accuracy is unclear.
\end{remark}

For the regularized ZF precoding, due to the complicated expressions in the precoding matrix and the power normalization coefficient, the calculations of the mean and variance of each random term in the SINR along with the CSI error noise power and power normalization coefficient are considerably more challenging, which are left to future work.

\section{Numerical Results}
\label{simulation}
In this section,  main analytical results in this paper will be verified by simulations. We assume that the total number of cells is $L=7$, the spatial correlation metric is $c=0.6,$ and the inter-cell normalized large scale fading is $\alpha=0.3$. The matrix $\bf A$ in \eqref{Eq4} and (\ref{Eq5}) is assumed to consist of the first $Mc$ columns of the $M$ by $M$ discrete fourier transfer (DFT) matrix. Due to the main focus of this paper, we only consider systems with non-decreasing SINR scenario as given in Corollary \ref{Corollary 1}. The first two subsections are for MRT and the last one is for ZF.

\subsection{MRT with Perfect Pilot Contamination Elimination}
First, we consider the case of perfect PCE, i.e., $L_{p}=0$ for all simulated antenna numbers. Eleven typical parameter settings as specified in Table I are considered.

\begin{table}[!hbp]\label{Table 1}
\centering
\caption{ Parameter settings for perfect PCE}
\begin{tabular}{|c|c|c|c|c|c|c|}
\hline
 & $E_t$ & $\rho$ & $K$ &  $r_s$& DE \\
\hline
Case 1 & 10 & 10 & $\left\lfloor {M/10 } \right\rfloor {\rm{ }}$ & 0&Y\\
\hline
Case 2 &10& $1/M$ & 10 &0&Y\\
\hline
Case 3 &10  &$1/\sqrt{M}$&$\left\lfloor {\sqrt M } \right\rfloor {\rm{ }}$  & 0&Y\\
\hline
Case 4 &10  &$1/\sqrt{M}$&$10$  & $0.5$&Y\\
\hline
Case 5 &10  &$10$&$\left\lfloor {\sqrt M } \right\rfloor {\rm{ }}$  & $0.5$&N\\
\hline
Case 6 &10  &$10$&$10$  & $1$&N\\
\hline
Case 7 &$10/M$& 10 & 10 &0&Y\\
\hline
Case 8 &$1/\sqrt{M}$  &$1/\sqrt{M}$  & 10 &0&Y\\
\hline
Case 9 &$1/\sqrt{M}$  &$10$  & $\left\lfloor {\sqrt M } \right\rfloor {\rm{ }}$ &0&Y\\
\hline
Case 10 &$1/\sqrt{M}$  &10  & 10 &0.5&Y\\
\hline
Case 11 &$10$  &$20/\sqrt{M}$  & 10 &0.5&*\\
\hline
\end{tabular}
\end{table}
In Table I, the $r_s$ value is the theoretical scaling exponent calculated by using $\eqref{Eq30}$. In the last column, ${\rm DE}=Y$ or $N$ indicates that the sufficient condition for the SINR to be asymptotically deterministic in \eqref{Eq33} is satisfied or not, respectively.

\begin{figure}[t]
\centering
\includegraphics[scale=0.5]{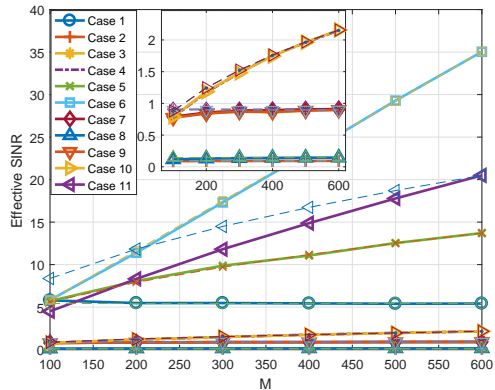}
\caption{Effective SINR of MRT for different parameter scaling with perfect PCE.}\label{Fig.4}
\end{figure}
In Fig. 1, the simulated effective SINR in (\ref{Eq24}) with respect to $M$ is shown for all parameter settings given in Table I to verify the performance scaling law in (\ref{Eq30}). The dotted lines are the corresponding reference curves with the theoretical scaling exponents. It can be shown that the scaling exponent of Case 6 is 1, the scaling exponents of Cases 4, 5 and 10 are 0.5, and the scaling exponents of Cases 1, 2, and 3, 7, and 8, 9 are 0, which is in accordance with the theoretical results given in Table I. The corresponding insights can be found in Discussions D1-D5.
On the other hand, the relatively high transmit power in Case 11 (although decreasing for larger $M$), makes the scaling law less accurate for the simulated large but finite antenna numbers. This is in accordance with Corollary \ref{corollary 2}. For Case 11, the condition in Corollary \ref{corollary 2} does not hold.

\begin{figure}[t]
\centering
\includegraphics[scale=0.5]{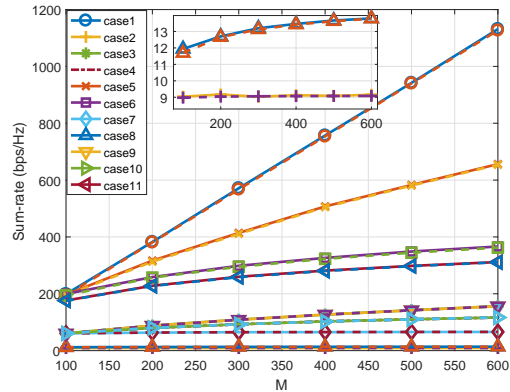}
\caption{Comparison between the sum-rate lower bound and simulated sum-rate with perfect PCE for MRT.}\label{Fig.4}
\end{figure}
In Fig.~2, the simulated achievable ergodic sum-rate for all parameter settings in Table I are compared with their corresponding lower bounds given in Lemma \ref{Lemma 2}. It can be shown that the lower bound is tight for all settings, which validates 1) the analysis based on it and 2) the effectiveness of the scaling law when the condition in Corollary \ref{corollary 2} is satisfied.

\begin{figure}[t]
\centering
\includegraphics[scale=0.5]{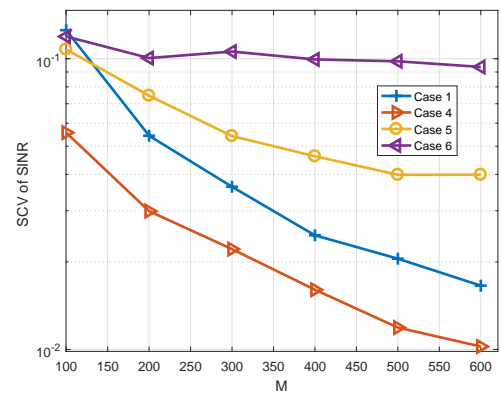}
\caption{The SCV of SINR of MRT for different parameter scaling with perfect PCE.}\label{Fig.4}
\end{figure}
In Fig. 3, the simulated SCVs of the SINRs are shown (in the logarithmic scale) for four of the typical cases. It can be shown that the SCVs for Cases 1 and 4 are smaller than those for Cases 5 and 6. Moreover, the decreasing exponents of the SCVs with respect to $M$ (fitted with the curve model $a/M^b$) for Cases 1 and 4 are about $1.1$ and $0.9$, respectively, while those for Cases 5 and 6 are about 0.6 and 0.12, respectively. These show that the SINRs of Cases 1 and 4 have faster decreasing SCVs with respect to $M$ than those of Cases 5 and 6, which is in accordance with the theoretical results given in Table I.

\subsection{MRT with Imperfect Pilot Contamination Elimination}
Then we turn to the case of imperfect PCE.
For space limit, two sets of values for $L_{p}$: $[5,5,5,5,5,5]$ and $[5,5,4,4,3,3]$, for the following antenna numbers: $[100,200,300,400,500,600]$, are considered. The scaling exponent of the former one is $r_\gamma=0$ corresponding to the case discussed in Section \ref{scalinglawres}. The scaling exponent of the latter one is approximated as $r_\gamma=0.35$ (via curve fitting) corresponding to the case discussed in Section \ref{decreasingPC}. The typical parameter settings are given in Table II.
\begin{table}[!hbp]\label{Table 2}
\centering
\caption{Network Parameters for Imperfect PCE}
\begin{tabular}{|c|c|c|c|c|c|}
\hline
& $E_t$ & $\rho$ & $K$ & $r_\gamma$ & $r_s$ \\
\hline
Case 1&0.2  & $1/\sqrt{M}$ &$\left\lfloor {\sqrt M } \right\rfloor {\rm{ }}$  &0.35 &0\\
\hline
Case 2&0.2  &0.1  & $\left\lfloor {M/10 } \right\rfloor {\rm{ }}$ &0.35 &0\\
\hline
Case 3&0.2   & 10 &2  &0.35 &0.35\\
\hline
Case 4&1  &$20$  & 2 &0 &0\\
\hline
Case 5&0.2  &$10$  & 10 &0 &0\\
\hline
\end{tabular}
\end{table}

\begin{figure}[t]
\centering
\includegraphics[scale=0.5]{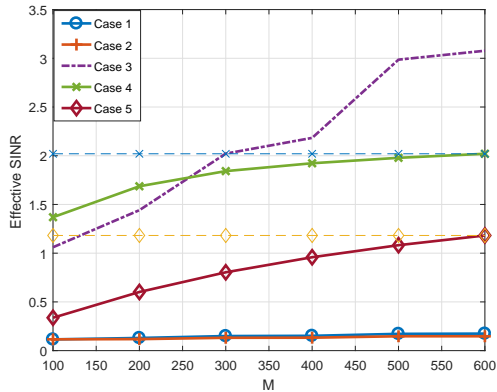}
\caption{Effective SINR of MRT for different parameter scaling with imperfect PCE.}\label{Fig.4}
\end{figure}
In Fig. 4, the simulated effective SINR with respect to $M$ is shown for the network settings given in Table II. For the constant pilot contamination $r_\gamma=0$ case, the limiting effect of pilot contamination on the performance scaling is shown in Cases 4 and 5. Note that these two cases do not match the scaling law as well as that of perfect PCE since the pilot contamination term is not dominant enough for the antenna number interval in both cases. However, for the simulated finite antenna numbers, the scaling law for Case 4 is more reliable compared with that for Case 5. This is because that the former case has a better compliance with the condition in Corollary \ref{corollary 3}. Further, as $M$ increases, the accuracy of the scaling law for both cases improves. On the other hand, Cases 1-3 show that for decreasing pilot contamination with respective to $M$,
an increasing SINR can be obtained (Case 3), which is in accordance with Remark \ref{Remark 6_170211}.

\begin{figure}[t]
\centering
\includegraphics[scale=0.5]{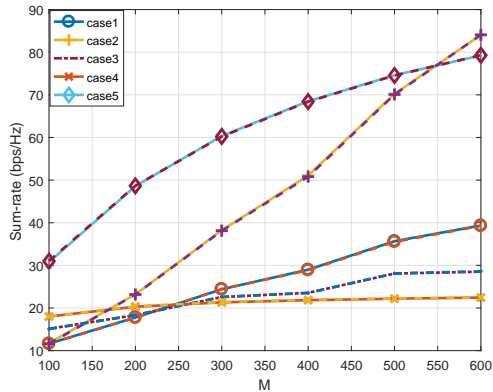}
\caption{Comparison between the sum-rate lower bound and simulated sum-rate value with imperfect PCE for MRT.}\label{Fig.4}
\end{figure}
In Fig.~5, the simulated sum-rate values and the lower bound given in Lemma \ref{Lemma 2} are shown for the network settings given in Table II. It can also been seen that the lower bound well matches the simulated values.


\subsection{Simulation Results for ZF}
The simulation results of ZF precoding for both perfect and imperfect PCE cases are provided in Fig. 6 and 7 to verify the scaling law. The parameter settings in Table I and II are used. It can be seen that the scaling exponents of the cases are basically the same as those of MRT. The only difference is that ZF has more relaxed SNR constraint for the applicability of the scaling law as noted in Remark \ref{zc_remark7}. This is the reason that the ZF scaling law result of Case 11 is also accurate.
\begin{figure}[t]
\centering
\includegraphics[scale=0.5]{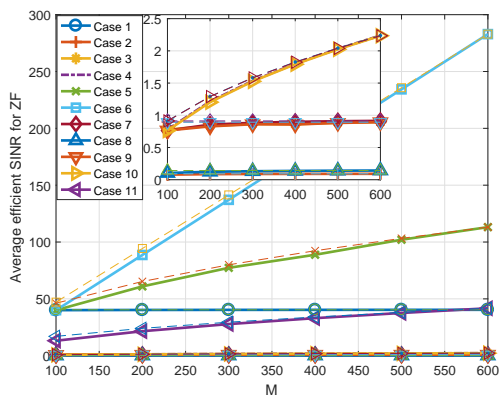}
\caption{Effective SINR of ZF for different parameter scaling with perfect PCE.}
\end{figure}
\begin{figure}[t]
\centering
\includegraphics[scale=0.5]{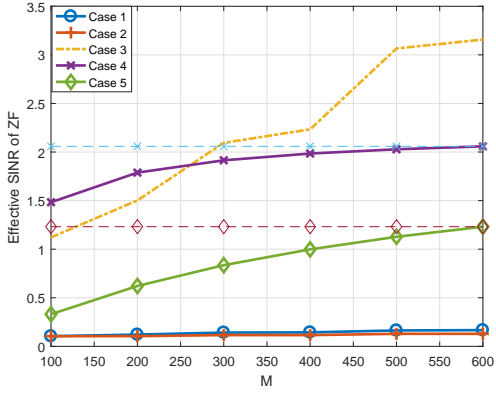}
\caption{Effective SINR of ZF for different parameter scaling with imperfect PCE.}
\end{figure}

\section{Conclusion}
For TDD mutli-cell multi-user massive MIMO downlink systems with CSI error, channel spatial correlation under MRT and ZF precodings, this paper provided a comprehensive general scaling law based performance analysis, which quantitatively shows the effect of important network parameters and the pilot contamination elimination and the tradeoff among them. The application of the scaling law for large but finite antenna number was also studied. Moveover, for MRT precoding, a sufficient condition for the SINR to be asymptotically deterministic in the sense of mean square convergence was derived. Our results cover existing ones on such analysis as special cases and show the effect of pilot contamination more explicitly. For the future work, more general spatial correlation distribution and more practical modeling on the effect of pilot contamination elimination should be studied.


%

\appendices

\section{Proof of Lemma \ref{Lemma1}}\label{Appendix A}
Due to the relationship in \eqref{Eq11}, the key step for calculating the means and SCVs of the random SINR components and $P_e$ is the calculation of ${{\rm E}\left\{ {{\bf{\hat h}}_{llk}^H{{{\bf{\hat h}}}_{llk}}} \right\}}$, ${{\rm E}\left\{ {{\bf{\hat h}}_{llk}^H{{{\bf{\hat h}}}_{llk}}{\bf{\hat h}}_{ssi}^H{{{\bf{\hat h}}}_{ssi}}} \right\}}$, and ${\rm E}\left\{ {{{\left| {{\bf{\hat h}}_{llm}^H{{{\bf{\hat h}}}_{llk}}} \right|}^2}} \right\}$, ${\rm E}\left\{ {{{\left| {{\bf{\hat h}}_{llm}^H{{{\bf{\hat h}}}_{llk}}} \right|}^2}{{\left| {{\bf{\hat h}}_{ssm}^H{{{\bf{\hat h}}}_{ssi}}} \right|}^2}} \right\}$.
The following results are derived for all $ l,s \in \{ j\}\cup {\mathbb{S}}_j^{p,d}$ and $k,i\in\{ 1,...,K\}$.

First, with
some tedious but straightforward calculations (the details of which are omitted here), we have
\begin{eqnarray}
&&\hspace{-7mm}{\rm E}\left\{ {{\bf{\hat h}}_{llk}^H{{{\bf{\hat h}}}_{llk}}} \right\} = {Q}M,\label{Eq38}\\
&& \hspace{-7mm}
{\rm E}\left\{ {{\bf{\hat h}}_{llk}^H{{{\bf{\hat h}}}_{llk}}{\bf{\hat h}}_{ssi}^H{{{\bf{\hat h}}}_{ssi}}} \right\}
\hspace{-1mm}=\hspace{-1mm}\left\{ \begin{array}{ll}
\hspace{-2mm}\left( \frac{Q}{c} \right)^2\hspace{-1.5mm}\Delta\hspace{-1mm} \left( \Delta  + 1 \right) & \hspace{-2mm}s = l,i = k\\
\hspace{-2mm}\left( \frac{Q}{c} \right)^2\Delta ^2 & \hspace{-2mm}\left(s,i\right) \ne \left(l,k \right)
\end{array} \right. \hspace{-5mm},
 \label{Eq39}\\
&&\label{Eq40} \hspace{-7mm}
{\rm E}\left\{ {{{\left| {{\bf{\hat h}}_{llm}^H{{{\bf{\hat h}}}_{llk}}} \right|}^2}} \right\} = \left\{ {\begin{array}{ll}
\hspace{-2mm}\left( \frac{Q}{c} \right)^2 \Delta \left( {\Delta  + 1} \right) & k = m\\
\hspace{-2mm}\left( \frac{Q}{c} \right)^2\Delta  & k \ne m
\end{array}} \right., \\
&&\label{Eq41}
\hspace{-7mm}
{\rm E}\left\{ {{{\left| {{\bf{\hat h}}_{llm}^H{{{\bf{\hat h}}}_{llk}}} \right|}^2}{{\left| {{\bf{\hat h}}_{ssm}^H{{{\bf{\hat h}}}_{ssi}}} \right|}^2}} \right\}=\left( \frac{Q}{c} \right)^4\times\nonumber\\
&&\hspace{-8mm}=\hspace{-1mm}\Delta \hspace{-1mm}\left\{ \begin{array}{ll}
\hspace{-2mm}\left( \Delta\hspace{-1mm}+\hspace{-1mm} 1 \right)
\hspace{-0.5mm}\left( \Delta  \hspace{-1mm}+\hspace{-1mm} 2 \right)
\hspace{-0.5mm}\left(\Delta \hspace{-1mm}+\hspace{-1mm}3\right) & \hspace{-2mm}s = l,i =k= m,\\
\hspace{-2mm}2\Delta  & \hspace{-2mm}s = l,i = k,k \ne m\\
\hspace{-2mm}\left(\Delta \hspace{-0.5mm}+\hspace{-0.5mm}1 \right)\left(\Delta  \hspace{-0.5mm}+\hspace{-0.5mm}2\right) & \hspace{-2mm}s = l,i \ne k,k = m \text{ or } i = m \\
\hspace{-2mm}\left(\Delta  + 1\right) & \hspace{-2mm}s = l,i \ne k,k \ne m,i \ne m\\
\hspace{-2mm}\Delta \left(\Delta  + 1 \right)^2 & \hspace{-2mm}s \ne l,i = k = m\\
\hspace{-2mm}\Delta & \hspace{-2mm}s \ne l,k \ne m,i \ne m\\
\hspace{-2mm}\Delta \left(\Delta  + 1\right) & \hspace{-2mm}s \ne l,i \ne k,k = m\text{ }{\rm{ or }}\text{ }i = m
\end{array} \right.\hspace{-6mm},
\end{eqnarray}
where \eqref{Eq38}-\eqref{Eq41} are derived from properties of Gamma distribution and in \eqref{Eq41}, the central limit theorem is used to approximate the distribution of $\frac{1}{\sqrt{M}}{\bf \hat h}_{llk}^H{\bf \hat h}_{ssi}$ as $\mathcal{CN}(0,Q^2/c)$ for all $(l,k)\neq(s,i)$ based on \eqref{Eq9}.

Based on (\ref{Eq38})-(\ref{Eq41}) and the relationship in (\ref{Eq11}), the next step for calculating the means and SCVs of the SINR components and $P_e$ is to follow their definitions and find the highest order terms with respect to $M$.

\section{PROOF OF PROPOSITION \ref{Pro1}}\label{Appendix B}
The SINR expression in (\ref{Eq17}) can be reformulated as
\begin{equation}\label{Eq42}
\begin{array}{l}
\hspace{-15pt}{\rm SINR}_{jm}{\rm{ = }}\frac{{{{{{\rm{M}}^{{r_s}}} \cdot {P_{\rm{s}}}} \mathord{\left/
 {\vphantom {{{{\rm{M}}^{{r_s}}} \cdot {P_{\rm{s}}}} {Q^2}}} \right.
 \kern-\nulldelimiterspace} {Q^2}}}}{{\frac{{\left( {K - 1} \right){P_{i,in}}}}{{Q^2{M^{1 - {r_s}}}}} + \frac{{MK{L_{p}}{P_{i,out}}}}{{Q^2{M^{1 - {r_s}}}}} + \frac{{K({L_{p}} + 1){P_e}}}{{Q^2{M^{1 - {r_s}}}}} + \frac{{K{Q}}}{{\rho Q^2{M^{1 - {r_s}}}}}}}
\end{array}.\hspace{-7pt}
\end{equation}
The SINR is asymptotically deterministic when its SCV approaches zero as ${M \to \infty }$. However, the complex structure of the SINR expression makes it challenging to obtain its SCV directly. Alternatively, since $P_{s}$ is asymptotically deterministic as noted in Remark \ref{Remark 1}, for the SINR to be asymptotically deterministic, the sufficient and necessary condition is that the denominator in (\ref{Eq42}) is asymptotically deterministic. One sufficient condition is the SCV of the denominator in (\ref{Eq42}) scales no larger than $1/M$. When ${\rm SINR}_{jm}$ is asymptotically deterministic, ${\rm SINR}_{jm}$ asymptotically converges to ${\rm {\widetilde {SINR}}}_{jm}$ as
defined in (\ref{Eq27}). Thus from the definition in (\ref{Eq28}), i.e., ${\rm {\widetilde {SINR}}}_{jm} = \mathcal{O}\left( {{M^{{r_s}}}} \right)$, and ${{{P_s}} \mathord{\left/
 {\vphantom {{{P_s}} {Q^2\mathop  \to \limits^{m.s} }}} \right.
 \kern-\nulldelimiterspace} {Q^2\mathop  \to \limits^{m.s.} }}1$ (refer to Lemma \ref{Lemma1}), we have
\begin{eqnarray}
&&\hspace{-7mm}{\rm E}\hspace{-0.5mm}\left\{ \hspace{-0.5mm}{\frac{{\left(\hspace{-0.5mm}  {K \hspace{-0.5mm}- \hspace{-0.5mm}1}\hspace{-0.5mm}  \right)\hspace{-1mm} {P_{i,in}}}}{{Q^2{M^{1 - {r_s}}}}} \hspace{-0.5mm}+\hspace{-0.5mm} \frac{{M\hspace{-0.5mm} K\hspace{-0.5mm} {L_{p}}\hspace{-0.5mm} {P_{i,out}}}}{{Q^2{M^{1 - {r_s}}}}} \hspace{-0.5mm}+\hspace{-0.5mm}
\frac{{K\hspace{-0.5mm} (\hspace{-0.5mm} {L_{p}} \hspace{-0.5mm}+\hspace{-0.5mm} 1)\hspace{-0.5mm} {P_e}}}{{Q^2{M^{1 - {r_s}}}}} \hspace{-0.5mm}+\hspace{-0.5mm} \frac{{K{Q}}}{{\rho Q^2{M^{1 - {r_s}}}}}} \hspace{-1mm} \right\} \nonumber\\
&&=\mathcal{O}\left( 1 \right).
\end{eqnarray}
Therefore, the sufficient condition can be rewritten as
\begin{equation}\label{Eq45}
{\mathop{\rm Var}} \left\{ \frac{{\left( {K - 1} \right){P_{i,in}}}}{{Q^2{M^{1 - {r_s}}}}} + \frac{{MK{L_{p}}{P_{i,out}}}}{{Q^2{M^{1 - {r_s}}}}}  \right\} \le \frac{{{C_1}}}{M}
\end{equation}for some constant ${C_1}$.
A sufficient condition for (\ref{Eq45}) is that the variance of each term in (\ref{Eq45}) scales no larger than $1/M$, i.e.,
\begin{eqnarray}
&&  {\mathop{\rm Var}} \left\{ {\frac{{\left( {K - 1} \right){P_{i,in}}}}{{Q^2{M^{1 - {r_s}}}}}} \right\} \le \frac{{{C_2}}}{M} \label{51} \\
&&  {\mathop{\rm Var}} \left\{ {\frac{{MK{L_{p}}{P_{i,out}}}}{{Q^2{M^{1 - {r_s}}}}}} \right\} \le \frac{{{C_3}}}{M} \label{52}
\end{eqnarray}
for some constant ${C_2}$ and ${C_3}$.

By combining the scaling law in (\ref{Eq28}-\ref{Eq30}) and the results in Lemma 1, the condition (\ref{51}) becomes
\setlength{\arraycolsep}{1pt}
\begin{eqnarray*}
&&\mathop {\lim }\limits_{M \to \infty } \frac{{\log \left\{ \left( {K - 1} \right)\left( {\frac{1}{{{M^{2 - 2{r_s}}}}}} \right)\frac{1}{{{c^2}}}\left( {\frac{{K - 1}}{{Mc}} + 1} \right)\right\}}}{{\log M}}\nonumber \\
&=&{r_k} - \left( {2 - 2{r_s}} \right) + \max \left( {{r_k} - 1,0} \right) \nonumber\\
&=&{r_k} - 2 + 2{r_s}\le \mathop {\lim }\limits_{M \to \infty } \frac{{\log \frac{{{C_2}}}{M}}}{{\log M}}= - 1,
\end{eqnarray*}
i.e.,
\begin{equation*}{r_k} + 2{r_s} \le 1.
\end{equation*}
Similarly, the condition (\ref{52}) becomes
\begin{equation*}
r_s\le 0 \quad\text{for}\quad L_p \ne 0.
\end{equation*}
Therefore, for perfect PCE, the sufficient condition is
\begin{equation*}
{r_k} + 2{r_s} \le 1 \quad\text{and}\quad {r_s} = 1 - {r_t} - {r_k} - {r_\rho},
\end{equation*}
which can be simplified to \eqref{Eq33}.
For imperfect PCE, the sufficient condition is
\begin{equation*}
{r_k} + 2{r_s} \le 1, r_s\le 0 \text{ and } {r_s}= \min \left( {1 - {r_t} - {r_k} - {r_\rho},0} \right),
\end{equation*}
which can be surely satisfied.

\section*{acknowledgement}
The authors would like to thank Associate Editor Prof. Cheng-Xiang Wang and the four anonymous reviewers for their helpful and insightful comments.



\ifCLASSOPTIONcaptionsoff
  \newpage
\fi

\begin{IEEEbiography}[{\includegraphics[width=1in,height=1.25in,clip,keepaspectratio]{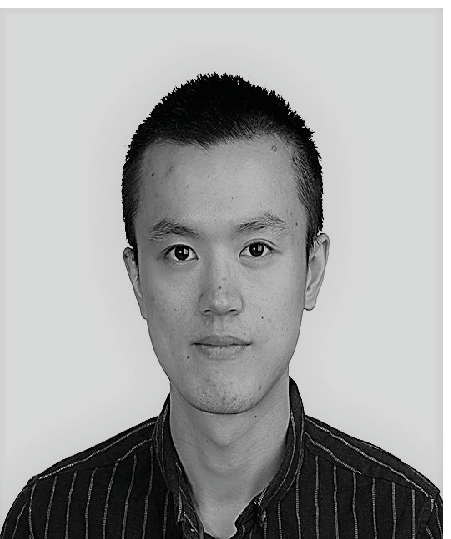}}]{Cheng Zhang}received the B.Eng. degree from Sichuan University, Chengdu, China in 2009, and the M.Sc. degree from Institute No.$206$ of China Arms Industry Group Corporation, Xian, China in 2012. Since March 2014, he has been working towards Ph.D. degree at Southeast University, Nanjing, China. His research interests include space-time signal processing and channel estimation in massive multi-input-multi-output and millimeter wave communication systems.
\end{IEEEbiography}
\begin{IEEEbiography}[{\includegraphics[width=1in,height=1.25in,clip,keepaspectratio]{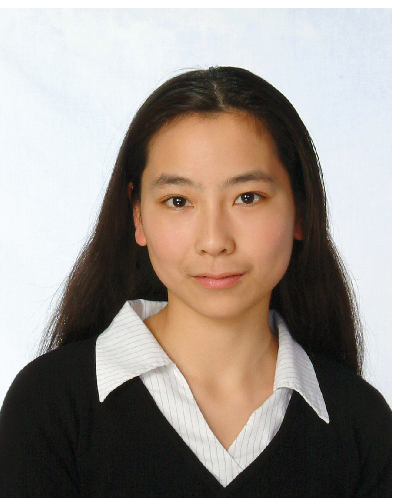}}]{Yindi Jing}received the B.Eng. and M.Eng. degrees from the University of Science and Technology of China, in 1996 and 1999, respectively. She received the M.Sc. degree and the Ph.D. in electrical engineering from California Institute of Technology, Pasadena, CA, in 2000 and 2004, respectively. From Oct. 2004 to Aug. 2005, she was a postdoctoral scholar at the Department of Electrical Engineering of California Institute of Technology. Since Feb. 2006 to Jun. 2008, she was a postdoctoral scholar at the Department of Electrical Engineering and Computer Science of the University of California, Irvine. In 2008, she joined the Electrical and Computer Engineering Department of the University of Alberta, where she is currently an associate professor.
She was an Associate Editor for the IEEE Transactions on Wireless Communications 2011-2016 and currently serves as a member of the IEEE Signal Processing Society Signal Processing for Communications and Networking (SPCOM) Technical Committee. Her research interests are in massive MIMO systems, cooperative relay networks, training and channel estimation, robust detection, and fault detection in power systems.
\end{IEEEbiography}
\begin{IEEEbiography}[{\includegraphics[width=1in,height=1.25in,clip,keepaspectratio]{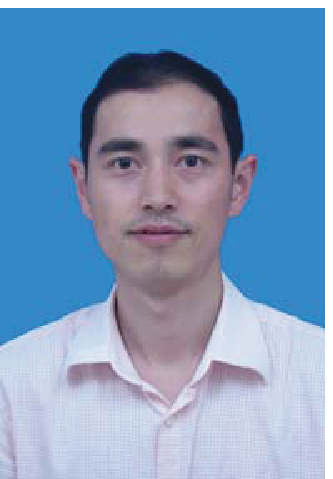}}]{Yongming Huang (M'10)} received the B.S. and M.S. degrees from Nanjing University, China, in 2000 and 2003, respectively, and the Ph.D. degree in electrical engineering from Southeast University, China, in 2007. Since Mar. 2007, he has been a faculty member with the School of Information Science and Engineering, Southeast University, where he is currently a full professor.
His current research interests include multiple-antenna wireless communications and signal processing.
He was an Associate Editor for the IEEE Transactions on Signal Processing 2012-2017. Since 2012, he has served as an Associate Editor for the EURASIP Journal on Advances in Signal Processing, and EURASIP Journal on Wireless Communications and Networking. He was selected as the Chang Jiang Young Scholar of Ministry of Education of China in 2015, and the recipient of the best paper awards from the 2012 and 2015 IEEE WCSP.

\end{IEEEbiography}
\begin{IEEEbiography}[{\includegraphics[width=1in,height=1.25in,clip,keepaspectratio]{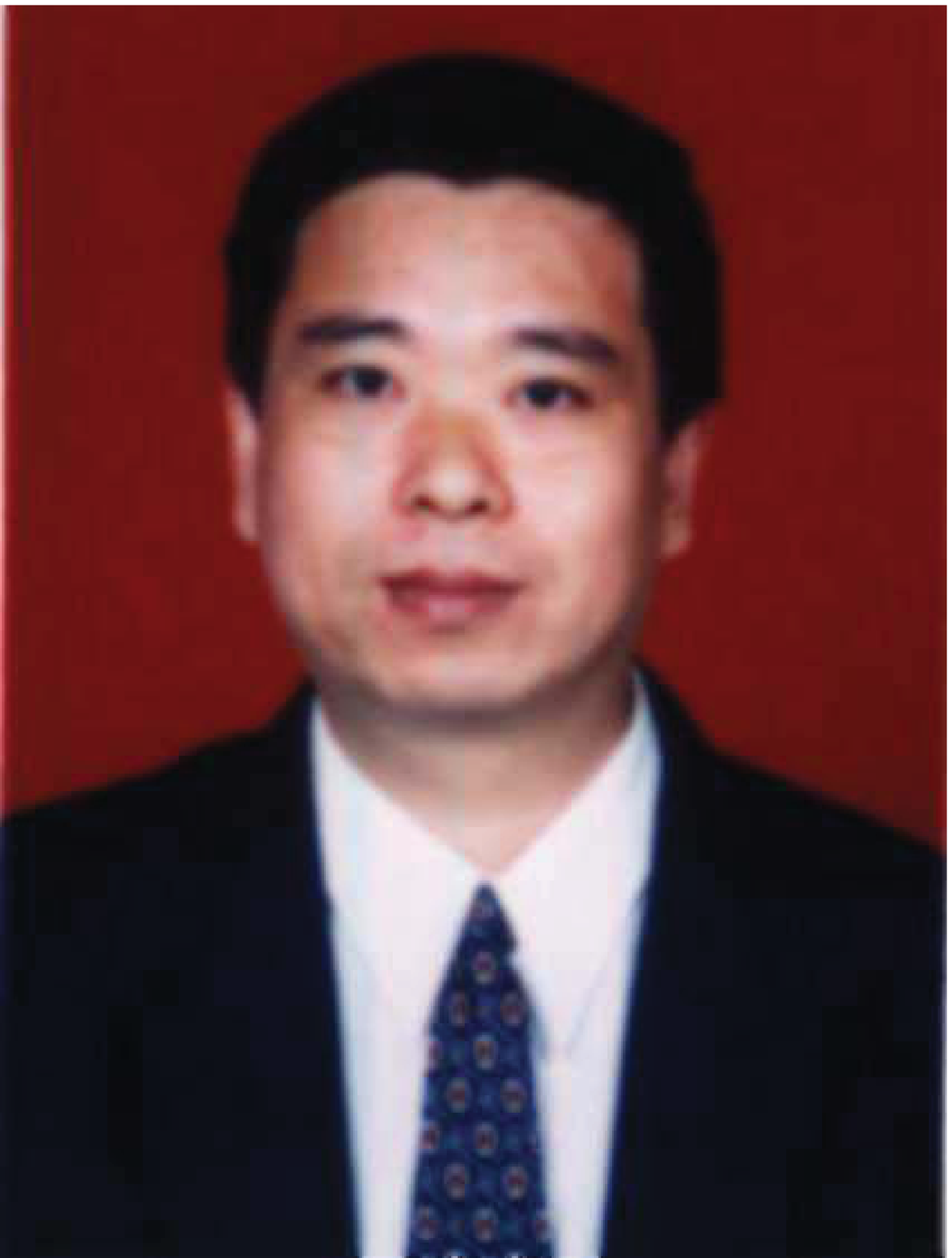}}]{Luxi Yang (M'96)} received the M.S. and Ph.D. degrees in electrical engineering from Southeast University, China, in 1990 and 1993, respectively. Since 1993, he has been with the Department of Radio Engineering, Southeast University, where he is currently a Professor of information systems and communications, and the Director of Digital Signal Processing Division. His current research interests include signal processing for wireless communications, MIMO communications, cooperative relaying systems, and statistical signal processing. He received the first- and second-class prizes of Science and Technology Progress Awards of the State Education Ministry of China in 1998 and 2002. He is currently a member of the Signal Processing Committee of Chinese Institute of Electronics.
\end{IEEEbiography}

\end{document}